\documentclass{aa}  

\usepackage{natbib}
\usepackage{graphicx}
\usepackage{txfonts}

\usepackage[dvipsnames]{xcolor}
\usepackage{tikz}
\usetikzlibrary{bayesnet}

\usepackage{diagbox}
\usepackage{subcaption}

\newcommand{\de}{\text{d}}
\newcommand{\Msun}{\text{M}_{\odot}}
\newcommand{\kmsec}{\text{km}\,\text{s}^{-1}}
\newcommand{\kmsecsq}{\text{km}^2\,\text{s}^{-2}}
\newcommand{\kpc}{\text{kpc}}
\newcommand{\pc}{\text{pc}}

\newcommand{\Msunppcc}{\Msun\pc^{-3}}
\newcommand{\Msunppcsquare}{\Msun\pc^{-2}}
\newcommand{\GeVcmcc}{\text{GeV}\,\text{cm}^{-3}}

\newcommand{\mas}{\text{mas}}

\newcommand{\popp}{\boldsymbol{\Psi}}

\newcommand{\data}{\boldsymbol{d}}

\newcommand{\sigm}{\text{sigm}}
\newcommand{\magn}{\text{mag}}

\defcitealias{PaperI}{Paper I}

\usepackage{hyperref}
\begin{document}

   \title{Weighing the Galactic disk using phase-space spirals \\ II. Most stringent constraints on a thin dark disk using Gaia EDR3}
   \titlerunning{Weighing the Galactic disk using phase-space spirals II}

   \author{A. Widmark
          \inst{1}
          \and
          C.F.P. Laporte
          \inst{2,3}
          \and
          P.F. de Salas
          \inst{4}
          \and
          G. Monari
          \inst{5}
          }

   \institute{Dark Cosmology Centre, Niels Bohr Institute, University of Copenhagen, Jagtvej 128, 2200 Copenhagen N, Denmark\\
   \email{axel.widmark@nbi.ku.dk}
   \and
   Institut de Ci\`encies del Cosmos (ICCUB), Universitat de Barcelona (IEEC-UB), Mart\'i i Franqu\`es 1, 08028 Barcelona, Spain
   \and
   Kavli Institute for the Physics and Mathematics of the Universe (WPI), The University of Tokyo Institutes for Advanced Study (UTIAS), The University of Tokyo, Chiba 277-8583, Japan
   \and
   The Oskar Klein Centre for Cosmoparticle Physics, Department of Physics, Stockholm University, AlbaNova, 10691 Stockholm, Sweden
   \and
   Universit\'e de Strasbourg, CNRS UMR 7550, Observatoire astronomique de Strasbourg, 11 rue de l'Universit\'e, 67000 Strasbourg, France
    }

   \date{Received Month XX, XXXX; accepted Month XX, XXXX}

 
  \abstract{Using the method that was developed in the first paper of this series, we measured the vertical gravitational potential of the Galactic disk from the time-varying structure of the phase-space spiral, using data from \emph{Gaia} as well as supplementary radial velocity information from legacy spectroscopic surveys. For eleven independent data samples, we inferred gravitational potentials that were in good agreement, despite the data samples' varied and substantial selection effects. Using a model for the baryonic matter densities, we inferred a local halo dark matter density of $0.0085 \pm 0.0039~\Msunppcc = 0.32 \pm 0.15~\GeVcmcc$. We were also able to place the most stringent constraint on the surface density of a thin dark disk with a scale height $\leq 50~\pc$, corresponding to an upper 95~\% confidence limit of roughly $5~\Msunppcsquare$ (compared to the previous limit of roughly $10~\Msunppcsquare$, given the same scale height). For the inferred halo dark matter density and thin dark disk surface density, the statistical uncertainties are dominated by the baryonic model, which potentially could also suffer from a significant systematic error. With this level of precision, our method is highly competitive with traditional methods that rely on the assumption of a steady state. In a general sense, this illustrates that time-varying dynamical structures are not solely obstacles to dynamical mass measurements, but they can also be regarded as assets containing useful information.}


   \maketitle
%

\section{Introduction}\label{sec:intro}

Our knowledge about the dynamics, composition, and history of the Milky Way is intimately connected to the determination of its gravitational potential \citep{1998MNRAS.294..429D,Klypin:2001xu,Widrow:2008yg,2014ApJ...794...59K,2017MNRAS.465..798C,2017MNRAS.465...76M,2020MNRAS.494.6001N,2020MNRAS.494.4291C,2020ApJ...894...10L}. Furthermore, direct and indirect dark matter detection experiments rely on precise knowledge on how dark matter is distributed in our Galaxy \citep{stoehr03,vogelsberger09,2015PrPNP..85....1K,2020PhRvD.102l3028P,Nobile:2021qsn}. The gravitational potential of the Galaxy is often inferred by fitting a stellar number density distribution to data under the assumption of a steady state, either in the solar neighbourhood \citep{2015ApJ...814...13M,WidmarkMonari,Sivertsson:2017rkp,Schutz:2017tfp,Buch:2018qdr,Guo:2020rcv,Salomon:2020eer,2021MNRAS.503.1586L} or a more global spatial volume (\citealt{2013JCAP...07..016N,2017MNRAS.465..798C,2017MNRAS.465...76M,2020MNRAS.494.4291C,Hattori:2020eft,2020PhRvD.102l3028P}). Other studies have explored a measurement of the gravitational potential directly from stellar accelerations \citep{Chakrabarti:2020abx,Buschmann:2021izy} or from stellar streams \citep{2010ApJ...712..260K,2018ApJ...867..101B,mahlan19,widmark_streams}.

Gravitational probes of the Milky Way can help constrain the particle nature of dark matter through the detection or exclusion of dark substructures, such as the subhalos predicted in the cold dark matter scenario \citep{diemand08, springel08, stref_lavalle17,facchinetti2020}. There has also been speculation that a dark disk, co-planar with the stellar disk, could exist in the Milky Way, formed either from the accretion of satellites \citep{10.1111/j.1365-2966.2008.13643.x,0004-637X-703-2-2275,2014MNRAS.444..515R} or, in a less standard scenario, from a dark matter particle sub-species with strong dissipative self-interactions \citep{Fan:2013tia,Fan:2013yva}. This latter type of dark disk could potentially be very thin, with a scale height as small as a few tens of parsecs. Such a disk has been constrained in dynamical mass measurements of the solar neighbourhood
\citep{Kramer:2016dqu,Caputo:2017zqh,Schutz:2017tfp,Buch:2018qdr}, where the strongest constraints use \emph{Gaia} observations of the very local population of stars. However, in a similar study by \cite{2021A&A...646A..67W} (also drawing from the results of \citealt{Widmark2019}), they argue that such measurements are strongly biased by time-varying dynamical effects.

Although disk formation requires a fairly quiescent accretion history \citep{freeman02}, non-equilibrium effects are certainly not lacking in the Milky Way. Over the last decades, a growing body of observations, partly enabled by the advent of large automated Milky Way surveys, have revealed signs of large-scale asymmetries in the vertical structure in the solar neighbourhood \citep{2012ApJ...750L..41W} and the outskirts of the Galactic disk \citep{newberg02}, as well as in its velocity distribution \citep{2012ApJ...750L..41W, williams13, carlin13}. This suggests the presence of bending waves in the disk \citep{xu15, price-whelan15, sheffield18, bergemann18}, possibly sourced by self-excitation \citep{chequers17}, satellite and dark subhalo interactions \citep{kazantzidis08,gomez13, widrow14, chequers18}, fly-bys \citep{gomez16}, or even bar buckling \citep{khoperskov19}. The \emph{Gaia} satellite reaffirmed the presence of these asymmetries and helped characterise the wavelength of these perturbations \citep[e.g.][]{schoenrich18}, as well as their time-evolving nature. Using the second data release from \emph{Gaia}, \cite{2018Natur.561..360A} revealed the presence of a phase-space spiral in the solar neighbourhood. The presence of the phase-space spiral at all stellar ages, as shown by \cite{laporte19}, indicates a recent collective perturbation of the disk. Furthermore, the spiral's presence over more than two disk scale lengths \citep{laporte19} and as far as $14~\kpc$ from the Galactic centre \citep{xu20} confirms it being a manifestation of a disk-wide disequilibrium phenomenon. This reinforces the connection between the perturbations seen in the solar neighbourhood and the Galactic outskirts, as was originally predicted by pre-\emph{Gaia} models of satellite interactions \citep{laporte18}, such as the Sagittarius dwarf galaxy. Indeed, using toy models, \cite{2018Natur.561..360A} and \cite{BS18} noted that the perturbation may have been set off as far back as 900 Myr ago, which is similar to the orbital period of the Sagittarius dwarf galaxy derived from stream fitting models \citep[e.g.][]{johnston05, vasiliev2021}.
Although the phase-space spiral was originally identified and mostly studied in its azimuthal and radial velocity moments (e.g. \citealt{2018Natur.561..360A, BS18, khoperskov19, bh19}), we exploit the spiral's shape as seen in terms of its relative number density with respect to the stellar bulk background, as first revealed in \cite{laporte19}. This reduces the exercise of potential fitting to only the vertical dimension, similar to what is typically done when determining the gravitational potential and dark matter density in the solar neighbourhood \citep{Read2014,2020arXiv201211477D}.

The phase-space spiral is a time-varying dynamical structure and as such it constitutes an obstacle and a systematic bias to traditional dynamical mass measurement methods that assume a steady state (e.g. Jeans modelling). The method employed in this work is complementary to such traditional methods, in the sense that it extracts information from the shape of the phase-space spiral itself. It does so under the assumption that the winding angle of the spiral is a smooth function with respect to vertical energy; given the shape of spiral in the $(z,w)$-plane, this sets strong constraints on the vertical gravitational potential. The general principles of our method are discussed at length in the first paper of this series---\cite{PaperI}, henceforth referred to as \citetalias{PaperI}---where we also tested our method on one-dimensional simulations. In those tests, we were able to retrieve the true gravitational potentials of our simulations with high accuracy.

In this work, we applied our method to the Milky Way, using the early instalment of \emph{Gaia}'s third data release (EDR3), supplemented with radial velocity measurements from legacy spectroscopic surveys through the catalogue compiled by \cite{SD18}. We constructed eleven main data samples using different cuts in Galactocentric radius and angular momentum. We were able to measure the gravitational potential to high precision and, using a model for the baryonic densities, we inferred the local halo dark matter density and placed the most stringent constraints on the surface density of a thin dark disk.

This article is structured as follows. We begin with some definitions in Sect.~\ref{sec:definitions}, such as the coordinate system, continuing with a description of the data in Sect.~\ref{sec:data}. In Sects.~\ref{sec:model} and \ref{sec:baryonic_model}, we outline our model of inference and our model for the baryonic matter densities in the solar neighbourhood. In Sect.~\ref{sec:results}, we present our results. Finally, we discuss and conclude in Sects.~\ref{sec:discussion} and \ref{sec:conclusion}.

\section{Coordinate system and other definitions}\label{sec:definitions}

In this paper, we use the following system of coordinates. The spatial coordinates $\boldsymbol{X} \equiv \{X,Y,Z\}$ denote the position with respect to the Sun, where positive $X$ corresponds to the direction of the Galactic centre, positive $Y$ corresponds to the direction of Galactic rotation, and positive $Z$ corresponds to the direction of Galactic north. Their respective time derivatives correspond to velocities $\boldsymbol{V} \equiv \{U,V,W\}$ in the solar rest frame.

The spatial coordinate $Z$, denoting height with respect to the Sun, is related to the height with respect to the Galactic disk, written $z$, according to
\begin{equation}
    z = Z + Z_\odot,
\end{equation}
where $Z_\odot$ is the Sun's height with respect to the Galactic mid-plane. Similarly, the velocities in the Local Standard of Rest \citep[see][]{BT2008}, written $\boldsymbol{v} \equiv \{u,v,w\}$, are found via the Sun's peculiar motion, written $\boldsymbol{V}_\odot \equiv \{U_\odot,V_\odot,W_\odot\}$, according to
\begin{equation}
    \boldsymbol{v} = \boldsymbol{V} + \boldsymbol{V}_\odot.
\end{equation}
We adopt values of $\boldsymbol{V}_\odot = \{11.1,12.24,7.25\}~\kmsec$ \citep{2010MNRAS.403.1829S}.

The Poisson equation is written
\begin{equation}\label{eq:Poisson}
	\frac{\partial^2\Phi(R,z)}{\partial z^2} + \frac{1}{R}\frac{\partial}{\partial R}
    \left[ R \frac{\partial\Phi(R,z)}{\partial R} \right] = 4\pi G \rho(z),
\end{equation}
where we have neglected the contribution in the azimuthal direction (which is zero in the case of rotational symmetry). The second term of the left-hand side is known as the circular velocity term, because the circular velocity of the Galactic plane $v_\text{c}$ is given by
\begin{equation}
    v^2_{\text{c}}(R) =  R \frac{\partial\Phi(R,0~\pc)}{\partial R}.
\end{equation}
This term can be written as a matter density correction, as
\begin{equation}\label{eq:rho_correction}
    \Delta \rho = \frac{1}{4\pi G R}\frac{\partial}{\partial R}
    \left[ R \frac{\partial\Phi(R,0~\pc)}{\partial R} \right].
\end{equation}
Because the rotational velocity curve is close to flat, this correction is small; the slope of the circular velocity is roughly $\partial v_{\rm{c}} / \partial R = -1.5\pm0.2~\kmsec\, \kpc^{-1}$ ($-1.7\pm 0.1~\kmsec\, \kpc^{-1}$ in \citealt{2019ApJ...871..120E}; $-1.33\pm 0.1~\kmsec\, \kpc^{-1}$ in \citealt{2020ApJ...895L..12A}), giving $\Delta \rho \simeq -0.0016\pm0.0006~\Msunppcc$.

In our model of inference, the gravitational potential is equal to
\begin{equation}\label{eq:phi}
    \Phi(z\,|\,\rho_h) = \sum_{h=1}^{4}
    \frac{4 \pi G \rho_h}{(2^{h-1} \times 100~\pc)^2}\log\Bigg[\cosh\Bigg(\dfrac{z}{2^{h-1} \times 100~\pc}\Bigg)\Bigg],
\end{equation}
where the free parameters $\rho_{h=\{1,2,3,4\}}$ are constrained to lie in the range $[0,0.2]~\Msunppcc$. Via the Poisson equation of Eq.~\eqref{eq:Poisson} and also accounting for the matter density correction of Eq.~\eqref{eq:rho_correction}, the total matter density is equal to
\begin{equation}\label{eq:rho_parametrisation}
    \rho(z) = \Bigg[\sum_{h=1}^{4} \rho_h \cosh^{-2}\Bigg(\dfrac{z}{2^{h-1} \times 100~\pc}\Bigg)\Bigg] + \Delta \rho.
\end{equation}
Using this functional form, the gravitational potential and matter density distribution are very free to vary in shape (although assumed to be symmetric, smooth, and strictly decreasing with $|z|$). They are not constrained by strong prior information about the baryonic matter density distribution in the Galactic disk, but are flexible enough to emulate such models (see e.g. \citealt{2015ApJ...814...13M} and \citealt{Schutz:2017tfp}).

\section{Data}\label{sec:data}

In this work we have used data from \emph{Gaia} EDR3, supplemented with radial velocity information compiled in \cite{SD18} using LAMOST DR3 \citep{Deng12}, GALAH DR2 \citep{buder18}, RAVE DR5 \citep{kunder17}, APOGEE DR14 \citep{abolfathi18}, SEGUE \citep{Yanny09}, and GES DR3 \citep{Gilmore12}. A supplementary radial velocity was used if a star in the \emph{Gaia} catalogue had missing radial velocity information or had a radial velocity uncertainty larger than $3~\kmsec$. If there was radial velocity information from several supplementary surveys, we used the value with the smallest associated uncertainty. Additionally, we excluded any star with discrepant radial velocity measurements, requiring less than a $2.5\sigma$ tension between all supplementary surveys that had a radial velocity uncertainty smaller than $5~\kmsec$; this amounted to removing roughly one per cent of the stars with supplementary radial velocity measurements. The stellar number counts of the radial velocity information taken from the respective surveys are listed in Table~\ref{tab:RVs}.

{\renewcommand{\arraystretch}{1.6}
\begin{table}[ht]
	\centering
	\caption{Stellar number counts of radial velocity information for our eleven main data samples.}
	\label{tab:RVs}
    \begin{tabular}{| l | r | r |}
        \hline
		Survey & Number count & Percentage \\
		\hline
		\emph{Gaia} & 943,742 & 91.61 \\
		LAMOST & 38,271 & 3.72 \\
		GALAH & 22,834 & 2.22 \\
		RAVE & 12,273 & 1.19 \\
		APOGEE & 11,530 & 1.12 \\
        SEGUE & 950 & 0.09 \\
        GES & 518 & 0.05 \\
		\hline
	\end{tabular}
\end{table}}

We constructed eleven main data samples on which we applied our method. To begin with, we made cuts in data quality, which all samples were subjected to. After that, we constructed our respective samples by making cuts in phase-space. These cuts are discussed in detail below.

\subsection{Data quality cuts}\label{sec:quality_cuts}

In this work, we applied quite strong constraints with respect to data quality. This was done in order to be able to neglect observational uncertainties and utilise the stars' six-dimensional phase-space information. At the same time, this gave rise to strong selection effects, most importantly with regards to radial velocity completeness, which has a strong spatial dependence. However, strong selection effects are not detrimental to our method, as long as the shape of the phase-space spiral is robustly extracted (this is discussed further in the beginning of Sect.~\ref{sec:bulk_and_spiral}).

We required that the stars in our data sets would have a \emph{Gaia} $G$-band magnitude smaller than $15~\magn$ (in \emph{Gaia} EDR3 there are only spurious stars with radial velocity information above $15~\magn$, see \citealt{2021A&A...649A...1G}), and a radial velocity uncertainty ($\sigma_\text{RV}$) smaller than $3~\kmsec$. For the radial velocity measurements coming from the \emph{Gaia} spectrograph, the uncertainties are typically around $0.3~\kmsec$ for bright stars and $1.8~\kmsec$ for dim stars \citep{rv_systematics}, with only a thin tail of stars with uncertainties larger than $3~\kmsec$; as such, our cut in radial velocity uncertainty only removed a small fraction of stars from our data samples.

In terms of the astrometric measurements, we required that the renormalised unit weight error (RUWE) would be smaller than 1.4 and that the parallax uncertainty ($\sigma_{\varpi}$) would be smaller than $0.05~\mas$. The first of these criteria is already very constraining, such that applying the subsequent second criteria removed only one in a thousand stars.

In summary, the data quality cuts can be written as follows:
\begin{equation}
\begin{split}
    (\text{i}) & \quad G<15~\magn, \\
    (\text{ii}) & \quad \sigma_\text{RV} < 3~\kmsec, \\
    (\text{iii}) & \quad \text{RUWE}<1.4, \\
    (\text{iv}) & \quad \sigma_{\varpi}<0.05~\mas.
\end{split}
\end{equation}

\subsection{Phase-space cuts}\label{sec:phase-space_cuts}

\begin{figure}
	\includegraphics[width=1.\columnwidth]{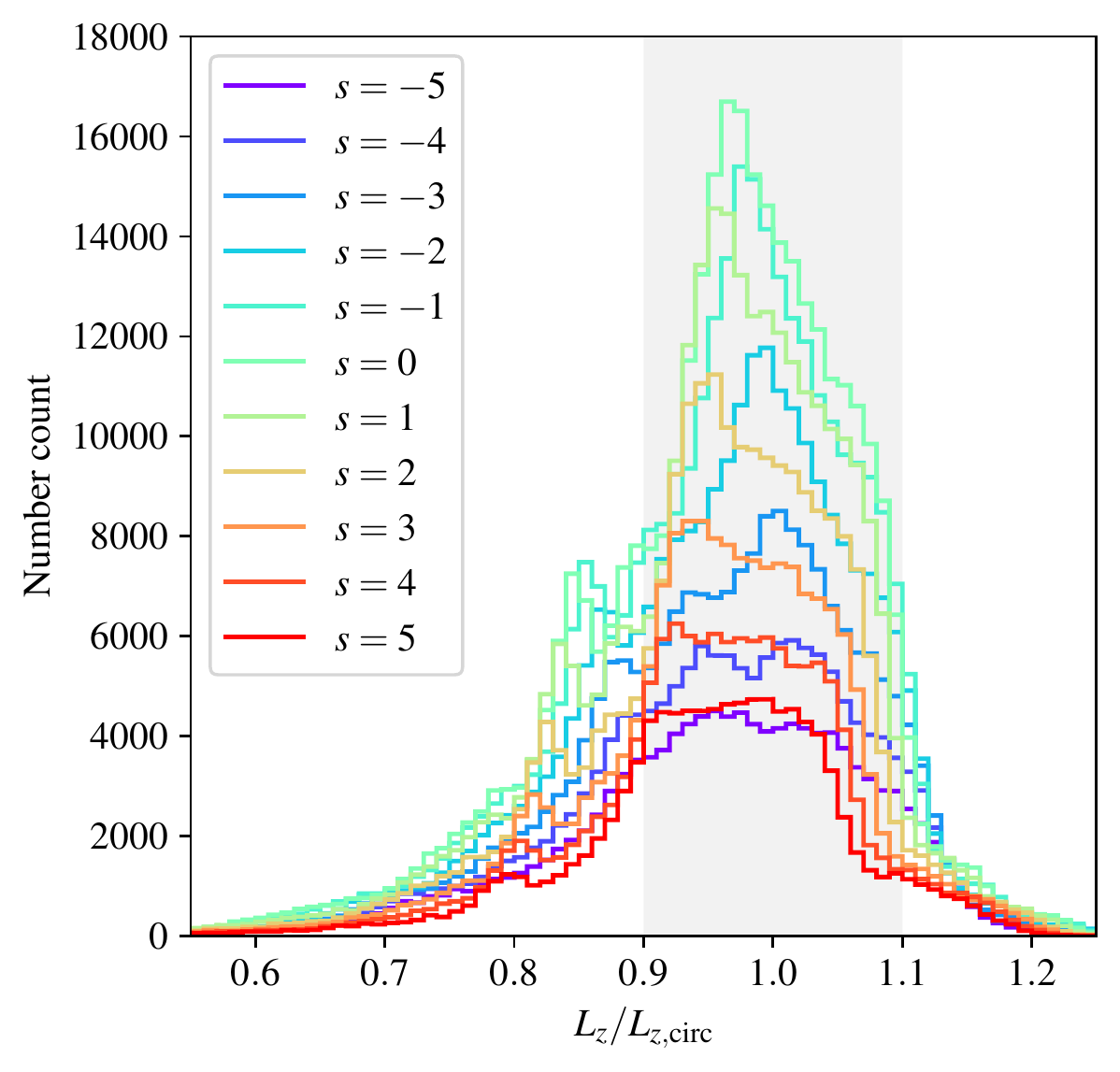}
    \caption{Histogram of angular momentum ($L_z$) relative to the angular momentum of a circular orbit ($L_{z,\text{circ.}}$), for different cuts in Galactocentric radius, where $R-R_\odot \in [100s-50, 100s+50]~\pc$. The grey band highlights the region where $L_z$ is within 10~\%  of $L_{z,\text{circ.}}$.}
    \label{fig:Lz_cuts}
\end{figure}

For the main data samples, we made the following phase-space cuts. We constructed bins in Galactocentric radius that were 100 pc wide, labelled by the index $s$, and also restricted the spatial extent in the azimuthal direction to $|Y| \leq 400~\pc$. In this spatial volume, we made cuts in angular momentum, requiring it to be close ($\pm 10~\%$) to that of a circular orbit. The angular momentum of a star is equal to
\begin{equation}
    L_z = v_\phi \times R,
\end{equation}
where $v_\phi$ is the velocity in the azimuthal direction in the Galactic rest frame. The angular momentum of a perfectly circular orbit was calculated by taking the data sample's Galactocentric mid-point and multiplying it by the circular velocity, according to
\begin{equation}
    L_{z,\text{circ.}} = v_{\rm{c}} \times [R_\odot+(100s~\pc)],
\end{equation}
assuming a flat rotation curve with a circular velocity of $v_{\rm{c}}=240~\kmsec$ \citep{2014ApJ...783..130R} and a solar Galactocentric radius of $R_\odot = 8178~\pc$ \citep{2019A&A...625L..10G}. In Figure~\ref{fig:Lz_cuts}, we show the histogram of the angular momentum relative to the angular momentum of a circular orbit. The $L_z$ distribution is not perfectly centred on $L_{z,\mathrm{circ}}$, but rather slightly shifted towards smaller values, as expected due to asymmetric drift \citep{BT2008}.

When making these cuts in data, we neglected any observational uncertainties associated with the measurements. For example, a star's coordinate $Y$ is taken directly from its angular position on the sky and its parallax value, according to $Y = \cos(l) \cos(b) \,(\mas/\varpi) \, \kpc$. This is motivated by our restrictive cuts in data quality; for example, with this parallax precision, the distance is known to a relative uncertainty of only a few percent at the furthest relevant distance of one kilo-parsec.

In summary, the phase-space cuts of the main data samples can be written:
\begin{equation}
\begin{split}
    (\text{i}) & \quad \frac{R-R_\odot}{\pc} \in [100s-50, 100s+50], \\
    (\text{ii}) & \quad \frac{Y}{\pc} \in [-400, 400], \\
    (\text{iii}) & \quad \frac{L_z}{v_c \times [R_\odot+(100s~\pc)]} \in [0.9,1.1],
\end{split}
\end{equation}
where $s$ is an integer in range -5 to 5.

To validate our choice of phase-space cuts, we also applied our method to a variety of data samples using different cuts. In particular, we include the results from an additional data sample where we only make a spatial cut according to $\sqrt{X^2+Y^2}<300~\pc$, with no cut in angular momentum. The total number of stars in our respective data samples is listed in Table~\ref{tab:number_counts}, counting only the stars for which $|Z|<800~\pc$.

{\renewcommand{\arraystretch}{1.6}
\begin{table}[ht]
	\centering
	\caption{Total number count of  of our main eleven data samples summed together, with the additional constraints that $|Z|<800~\pc$.}
	\label{tab:number_counts}
    \begin{tabular}{| l | r |}
        \hline
		Data sample & Number count \\
		\hline
		$s=-5$ & 36,810 \\
		$s=-4$ & 54,099 \\
		$s=-3$ & 78,581 \\
		$s=-2$ & 111,918 \\
		$s=-1$ & 150,048 \\
        $s=0$ & 171,432 \\
        $s=1$ & 145,883 \\
        $s=2$ & 108,238 \\
        $s=3$ & 76,247 \\
        $s=4$ & 52,402 \\
        $s=5$ & 35,727 \\
        $<300~\pc$ & 812,250 \\
		\hline
	\end{tabular}
\end{table}}

\subsection{Data reduction}\label{sec:reduction}

In our model of inference, the data was in the reduced form of a two-dimensional histogram in the $(Z,W)$-plane. This histogram is the number count of observed stars in bins of size $(20~\pc)\times(1~\kmsec)$, written $\data_{i,j}$, where the indices $(i,j)$ label the respective bins. Similar to the phase-space cuts defined in Sect.~\ref{sec:phase-space_cuts}, the values for $Z$ and $W$ are taken directly from the astrometric measurements, neglecting any observational uncertainties.

\section{Model of inference}\label{sec:model}

The method used in this work is the same as the one presented in \citetalias{PaperI}, although with some minor modifications, which are as follows: (i) we constrain ourselves to an asymmetric single-armed spiral; (ii) the spiral is fitted to slightly lower vertical energies; (iii) we mask the data for latitudes lower than $|b| \lesssim 20^{\circ}$; (iv) the Sun's vertical position and vertical velocity are fixed rather than free parameters. Our method is described below, with emphasis on its modifications with respect to the original version.

Just as for the tests run in \citetalias{PaperI}, we assume separability of the gravitational potential and reduce the dynamics to only the vertical dimension. In Jeans analysis, the so-called tilt term accounts for the coupling between the radial and vertical directions, and is a derivative of the bulk phase-space density distribution with respect to the radial direction. Because our method disregards the bulk density, no such term enters our method. The radial-vertical coupling of the gravitational potential does induce a radial motion, but for the relevant range of vertical energies ($E_z<\Phi(700~\pc)$) this motion is small compared to the radial extent of our data samples. In principle, this can still become relevant to the extent that the shape of the spiral has a radial dependence. Because the inferred gravitational potentials and phase-space spiral shapes are so similar between neighbouring data samples (see Sect.~\ref{sec:results} for more details), we deem such effects to be negligible, which motivates the assumption of vertical separability when using this method.

\subsection{Bulk and spiral phase-space densities}\label{sec:bulk_and_spiral}

In our method, we fit a phase-space distribution to data, consisting of a product of bulk and a spiral phase-space density distribution. In traditional methods that are based on the assumption of a steady state, the bulk density is the quantity that is used to infer the gravitational potential. Conversely, in our method the bulk is only fitted in order to extract the spiral shape and does not influence, nor is influenced by, the inferred gravitational potential. Because the bulk is fitted as a mere background, to a large extent it absorbs any selection effects pertinent to the data. Such effects are critical to account for in steady state modelling, but can be disregarded in our method as long as the shape of the spiral is robustly extracted.

The free parameters of our model are shown in Table~\ref{tab:model_parameters}. The parameters are split into two groups: the bulk phase-space density parameters, written $\popp_\text{bulk}$, and the spiral phase-space density parameters, written $\popp_\text{spiral}$. The total number of free parameters is equal to $7+3K$, where $K$ is the number of Gaussian components in the bulk stellar density, for which we set $K=6$.

{\renewcommand{\arraystretch}{1.6}
\begin{table}[ht]
	\centering
	\caption{Free parameters in our model of inference.}
	\label{tab:model_parameters}
    \begin{tabular}{| l | l |}
		\hline
		$\popp_\text{bulk}$  & Bulk phase-space density parameters \\
		\hline
		$a_k$ & Weights of the Gaussian mixture model \\
		$\sigma_{z,k}$, $\sigma_{w,k}$ & Dispersions of the Gaussian mixture model \\
		\hline
		\hline
		$\popp_\text{spiral}$  & Spiral phase-space density parameters \\
		\hline
		$\rho_{h=\{1,2,3,4\}}$ & Mid-plane matter densities \\
		$t$ & Time since the perturbation was produced \\
		$\tilde{\varphi}_0$ & Initial angle of the perturbation \\
		$\alpha$ & Relative density amplitude of the spiral \\
		\hline
	\end{tabular}
\end{table}}

We model the bulk density as a Gaussians mixture model according to
\begin{equation}\label{eq:bulk_density}
    B(z,w\,|\,\popp_\text{bulk}) =
    \sum_{k=1}^{K} a_k \,
    \dfrac{\exp\Bigg(-\dfrac{z^2}{2\sigma_{z,k}^2}\Bigg)}{\sqrt{2\pi\sigma_{z,k}^2}} \,
    \dfrac{\exp\Bigg(-\dfrac{w^2}{2\sigma_{w,k}^2}\Bigg)}{\sqrt{2\pi\sigma_{w,k}^2}} ,
\end{equation}
where $\popp_\text{bulk}$ are the bulk density components, which includes the Gaussian weights ($a_k$) and dispersions ($\sigma_{z,k}$, $\sigma_{w,k}$).

A star's vertical oscillation has a total time period of
\begin{equation}\label{eq:period}
    P(E_z\,|\, \rho_h) = \oint \frac{\de z}{w} =  4\int_0^{z_\text{max}} \frac{\de z}{\sqrt{2[E_z-\Phi(z \,|\, \rho_h)]}},
\end{equation}
where $E_z = \Phi(z \,|\, \rho_h) + w^2 / 2$ is a star's vertical energy per mass, $z_\text{max}$ is the maximum height that a star reaches, and $\rho_h$ parametrises the gravitational potential according to Eq.~\eqref{eq:phi}. The position of a star in the $(z,w)$-plane is associated with a temporal angle, which is given by
\begin{equation}\label{eq:angle_of_z}
\begin{split}
& \varphi(z,w \,|\, \rho_h) = \\
& \begin{cases}
    2 \pi P^{-1}{\displaystyle\int_0^{|z|}} \dfrac{\de z'}{\sqrt{2[E_z-\Phi(z' \,|\, \rho_h)]}} & \text{if}\,z\geq0\,\text{and}\,w\geq0, \\
    \pi - 2 \pi P^{-1}{\displaystyle\int_0^{|z|}} \dfrac{\de z'}{\sqrt{2[E_z-\Phi(z' \,|\, \rho_h)]}} & \text{if}\,z\geq0\,\text{and}\,w<0, \\
    \pi + 2 \pi P^{-1}{\displaystyle\int_0^{|z|}} \dfrac{\de z'}{\sqrt{2[E_z-\Phi(z' \,|\, \rho_h)]}} & \text{if}\,z<0\,\text{and}\,w<0, \\
    2\pi - 2 \pi P^{-1}{\displaystyle\int_0^{|z|}} \dfrac{\de z'}{\sqrt{2[E_z-\Phi(z' \,|\, \rho_h)]}} & \text{if}\,z<0\,\text{and}\,w\geq0.
\end{cases}
\end{split}
\end{equation}

In our analytic spiral model, the initial perturbation is assumed to have no initial winding. Because the gravitational potential is anharmonic, this perturbation winds into a spiral with time. Even though the initial perturbation might have a more complicated form, the winding behaviour can be expected to dominate the shape of the eventual spiral, as long as the perturbation is not a spiral-resembling shape to begin with (a more thorough discussion on the underlying assumptions of our spiral model can be found in \citetalias{PaperI}).
The spiral angle as a function of vertical energy $E_z$ evolves according to
\begin{equation}\label{eq:angle_of_time}
    \tilde{\varphi}(t,E_z \,|\, \rho_h,\tilde{\varphi}_0) = \tilde{\varphi}_0 + 2\pi\frac{t}{P(E_z \,|\, \rho_h)},
\end{equation}
where $\tilde{\varphi}_0$ is the initial angle of the perturbation.

The total phase-space density of our analytical model is equal to
\begin{equation}\label{eq:total_density}
    f(z,w\,|\,\popp) = B(z,w\,|\,\popp_\text{bulk})
    \times \Big[ 1 + m(z,w \, | \, \rho_h)\, S(z,w\,|\,\popp_\text{spiral}) \Big].
\end{equation}
In this expression,
\begin{equation}\label{eq:spiral_rel_density}
    S(z,w\,|\,\popp_\text{spiral}) =
    \alpha \cos\Big[ \varphi(z,w\,|\, \rho_h)-\tilde{\varphi}(t,E_z \,|\, \rho_h,\tilde{\varphi}_0) \Big],
\end{equation}
is the relative number density of the spiral with respect to the bulk, where $\alpha$ is a unit-less amplitude in range $[0,1]$. This is the first difference with respect to how the method was formulated in \citetalias{PaperI}, which included a symmetric spiral component with an amplitude $\beta$. This component was not included here, as no second arm is seen for the Milky Way spiral. Furthermore, Eq.~\eqref{eq:total_density} contains the quantity
\begin{equation}\label{eq:inner_boundary}
    m(z,w \, | \, \rho_h) = \text{sigm} \Bigg[
    \frac{E_z(z,w \, | \, \rho_h)-\Phi(300~\pc \, | \, \rho_h)}{\Phi(300~\pc \, | \, \rho_h)-\Phi(280~\pc \, | \, \rho_h)} \Bigg],
\end{equation}
where
\begin{equation}\label{eq:sigmoid}
    \sigm(x) \equiv \frac{1}{1+\exp(-x)}.
\end{equation} 
This sets a lower boundary in $E_z$ to the spiral of our analytic model; close to the origin of the $(z,w)$-plane, the spiral is washed out due to self-gravity effects. The numerical values of the $m(z,w \, | \, \rho_h)$ function constitute our second modification with respect to the method as formulated in \citetalias{PaperI}, in that the inner boundary is somewhat less restrictive, using a limit of $E_z \gtrsim \Phi(300~\pc)$ instead of $E_z \gtrsim \Phi(400~\pc)$. The reason for this change is that the phase-space spiral of the actual Milky Way is more clearly defined in the inner region than for the one-dimensional simulations of \citetalias{PaperI}, possibly because the effects of self-gravity are not as strong and cohesive for the more complex, three-dimensional kinematics of our Galaxy.

\subsection{Data likelihood and masks}\label{sec:likelihood_and_masks}

In our method, the phase-space density model was fitted to data in two separate steps. In the first step, we fitted the bulk density distribution without the spiral; in other words, we minimised the data likelihood with respect to $\popp_\text{bulk}$ while $\alpha=0$. In the second step, we fitted the relative phase-space density spiral; in other words, we minimised the likelihood with respect to $\popp_\text{spiral}$ while $\popp_\text{bulk}$ remained fixed.

The data likelihood is given by the Poisson count comparison of the model and data in the $(Z,W)$-plane, in bins labelled by the indices $(i,j)$. The logarithm of the likelihood is equal to
\begin{equation}\label{eq:likelihood}
\begin{split}
    & \ln\, \mathcal{L}(\data_{i,j}\,|\,\popp) = \\
    & - \sum_{i,j}
    \theta (|Z_i|-\bar{Z}) \, M(Z_i,W_j) \, \dfrac{[\data_{i,j}-f(Z_i+Z_\odot,W_j+W_\odot,\popp)]^2}{2 f(Z_i+Z_\odot,W_j+W_\odot\,|\,\popp)} \\
    & + \{\text{constant term}\},
\end{split}
\end{equation}
where $\theta (|Z_i|-\bar{Z})$ and $M(Z_i,W_j)$ are mask functions described below, and $f(Z_i+Z_\odot,W_j+W_\odot\,|\,\popp)$ is the model phase-space density as defined in Eq.~\eqref{eq:total_density}.

The first mask function in Eq.~\eqref{eq:likelihood} is a Heaviside step function equal to
\begin{equation}\label{eq:zmask}
    \theta (|Z_i|-\bar{Z}) =
    \begin{cases}
    0, & \text{if }|Z_i|\leq \bar{Z}, \\ 
    1, & \text{if }|Z_i|> \bar{Z}, \\ 
    \end{cases}
\end{equation}
where $\bar{Z} = \sin(20^{\circ}) \times |100s|~\pc$. This mask is applied to stars with low heights, roughly corresponding to a cut in Galactic latitude of $|b|<20^\circ$. This constitutes the third modification with respect to how the method was formulated in \citetalias{PaperI}. It is included in this work due to severe selection effects, especially in terms of the availability of radial velocity information close to the Galactic plane (see Sect.~\ref{sec:results} and Fig.~\ref{fig:spirals}).

The second mask function in Eq.~\eqref{eq:likelihood} is $M(Z,W)$, defining a circular outer boundary in the $(Z,W)$-plane. It is applied in order to mask the high vertical energies where the stellar number density is low and the spiral is less pronounced. It is defined
\begin{equation}\label{eq:mask}
    M(Z,W) =
    \text{sigm} \Bigg\{ -10\, \Bigg[\Bigg(\frac{Z}{Z_\text{lim.}}\Bigg)^2 + \Bigg(\frac{W+W_\odot}{W_\text{lim.}}\Bigg)^2 - 1 \Bigg] \Bigg\},
\end{equation}
where $\text{sigm}$ is the sigmoid function of Eq.~\eqref{eq:sigmoid}. The numerical values of $Z_\text{lim.}$ and $W_\text{lim.}$ differed for the two minimisation steps of our method, in order to avoid fitting artefacts. In the first step, when the bulk was fitted, we set $Z_\text{lim.}=800~\pc$ and $W_\text{lim.}=44~\kmsec$. In the second step, when the spiral was fitted, we set $Z_\text{lim.}=700~\pc$ and $W_\text{lim.}=40~\kmsec$.

The fourth and final modification of our method with respect to \citetalias{PaperI} is that we fixed the Sun's vertical position and velocity ($Z_\odot$ and $W_\odot$), rather than letting them be free parameters in our fitting procedure. This is motivated by significant and spatially dependent selection effects, coming from radial velocity completeness and other data quality cuts; completeness is poor close to the Galactic mid-plane and also has a strong asymmetry with respect to the Galactic north and south for many data samples. These issues made our inference of $Z_\odot$ and $W_\odot$ highly unstable and they are better determined in other studies. For the Sun's vertical velocity, we set $W_\odot=7.25~\kmsec$, which is a well established value, known to within an uncertainty smaller than $0.05~\kmsec$ \citep{2010MNRAS.403.1829S}. Within this uncertainty, varying $W_\odot$ has a negligible effect on our results. For the Sun's vertical position, the uncertainty is more significant. Different studies have produced somewhat discrepant results, ranging from roughly $0$--$20~\pc$ \citep{Juric:2005zr,2017MNRAS.468.3289Y,BovyAssym,Buch:2018qdr,2021A&A...646A..67W,2021A&A...649A...6G}. For this reason, we adopt three different values of $Z_\odot = \{0,10,20\}~\pc$ when applying our method, which allows us to estimate the uncertainty of our result with respect to the Sun's position.

Our method was implemented in \textsc{TensorFlow}, allowing for efficient minimisation using the Adam optimiser \citep{adamopt}. Still, minimising the spiral likelihood function is computationally expensive, requiring several hundred CPU hours. For a more detailed explanation of our method, as well as illustrative examples highlighting its general principles, we refer back to \citetalias{PaperI}. The code used in this work is open source and available online.\footnote{\url{https://github.com/AxelWidmark/SpiralWeighing}}

\subsection{Jackknifing}\label{sec:jackknife}

In order to estimate the statistical uncertainty of our respective data samples, we employed the technique of ``delete-$d$ jackknifing'' \citep{efron1982jackknife}. With this technique, the statistical uncertainty of some inferred parameter (e.g. $\psi$) is evaluated by inferring this parameter for a number of subsamples constructed from the full data sample. The statistical variance, written $\text{Var}(\psi)$, is proportional to the variance of the inferred parameter between the different subsamples, written $\text{Var}^*(\psi)$, according to
\begin{equation}
	\text{Var}(\psi) = \frac{n-d}{d} \text{Var}^*(\psi),
\end{equation}
where $n$ is the total number of data objects and $d$ is the number of deleted data objects.

Jackknifing should ideally be performed by running inference on all possible subsample constructions. This was not possible for us due to the computational cost of our algorithm. For each data sample and fixed value of $Z_\odot$, we independently constructed ten subsamples by randomly selecting half of the stars (i.e. $d=n/2$).

\section{Baryonic model}\label{sec:baryonic_model}

The gravitational potential of our model of inference, as stated in Eq.~\eqref{eq:phi}, did not rely on any baryonic model. However, after applying our method, we compared our results with a model for the baryonic matter densities, in order to constrain the halo dark matter density and the surface density of a thin dark disk.

We used the baryonic model from \cite{Schutz:2017tfp}, based on the pre-\emph{Gaia} studies by \cite{Flynn:2006tm}, \cite{2015ApJ...814...13M}, and \cite{0004-637X-829-2-126}. This baryonic model is also used in other local dynamical mass measurements, such as \cite{Sivertsson:2017rkp}, \cite{Buch:2018qdr}, \cite{Widmark2019}, and \cite{2021A&A...646A..67W}. In this model, the total baryonic density is a sum of twelve different components: four gas components (molecular, cold atomic, warm atomic, and hot ionised gas); six stellar components divided by absolute magnitude; white dwarfs; and brown dwarfs. Each of the twelve components is described in terms of its mid-plane matter density ($\rho_{t,0}$) and its vertical velocity dispersion ($\sigma_{w,t}$). It is assumed that the respective components are iso-thermal, such that their matter density profiles decay with height according to
\begin{equation}
    \rho_t(z) =
    \rho_{t,0}\exp\Bigg[ -\frac{\Phi(z)}{\sigma_{w,t}^2} \Bigg].
\end{equation}

\begin{figure*}
	\includegraphics[width=1.\textwidth]{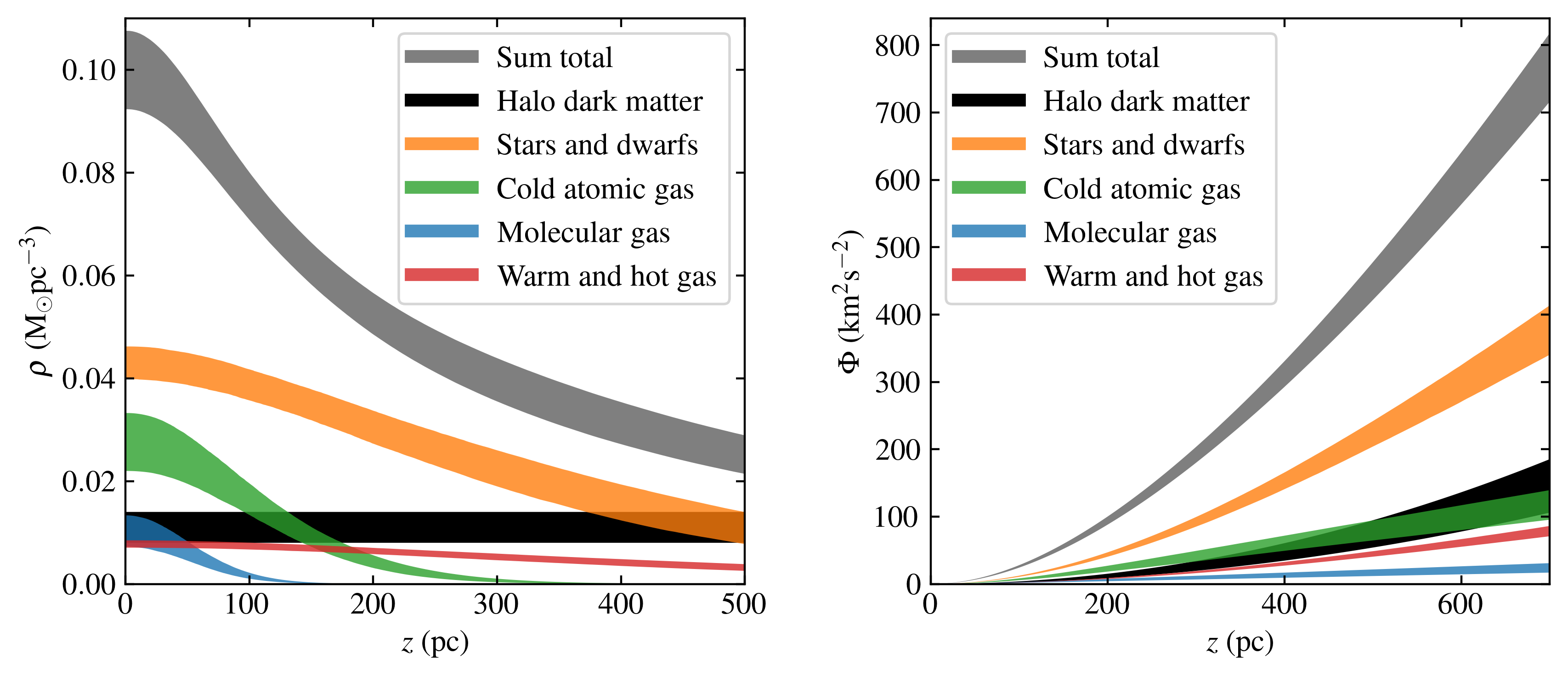}
    \caption{Baryonic model from \cite{Schutz:2017tfp}, including a component of halo dark matter, shown in terms of its matter density distribution (left panel) and vertical gravitational potential (right panel). Details are found in Sect.~\ref{sec:baryonic_model}.}
    \label{fig:baryons}
\end{figure*}

In Fig.~\ref{fig:baryons}, we show the matter density distribution and vertical gravitational potential of this baryonic model. In this figure, the baryons are divided into four different groups: stars and dwarfs, cold atomic gas, molecular gas, warm and hot gas. We also include a halo dark matter density of $0.011\pm0.03~\Msunppcc$, which encapsulates the approximate range of recent local dark matter density measurements \citep{2020arXiv201211477D}.

It is important to note that this model of baryonic matter densities in the solar neighbourhood could potentially suffer from significant systematic errors. For the stellar components, the assumption of iso-thermality are not ideal descriptions when comparing with \emph{Gaia} data: the stellar components' vertical velocity distributions are not strictly Gaussian but have heavier tails and the shape of the stellar number density profiles differ from those predicted by the baryonic model (a comparison can be found in the appendix of \citealt{2021A&A...646A..67W}). This is, at least in part, connected to the fact that the stellar components are categorised in terms of absolute magnitude, which does not differentiate them in terms of age and kinematic properties. Perhaps even more worrisome are the matter density distributions of the gas components, which have larger statistical uncertainties and are arguably more prone to significant systematic bias. For example, measuring molecular hydrogen depends on the CO-to-H$_2$ conversion factor, while measuring atomic hydrogen depends on corrections for optical depth (e.g. discussed in \citealt{2015A&A...579A.123H}). Furthermore, the cold gas is non-uniformly distributed, both in terms of density and ionisation \citep{2003A&A...411..447L}. As such, it is not at all impossible that the baryonic model suffers from systematic errors significantly larger than the reported statistical uncertainties.

\section{Results}\label{sec:results}

In this section, we present the results from having applied our method of inference on our respective data samples. Using a summary of those results, we also place constraints on the local dark matter density and the surface density of a thin dark disk. Most figures in this section only show a smaller number of data samples. The corresponding figures for the remaining data samples are found in Appendix~\ref{app:more_plots}.

In Fig.~\ref{fig:hists}, we show the data histograms of four representative data samples ($s=\{-4,-1,0,3\}$). The stellar number density differs quite dramatically between data samples (their number counts are listed in Table~\ref{tab:number_counts}), mainly depending on the distance. They vary in terms of their dispersion in $Z$, where the more distant data samples are wider while nearby data samples seem pinched (seen most clearly when comparing data samples $s=-4$ and $s=0$). Furthermore, the more distant data samples have a number density depression close to the Galactic mid-plane (i.e. for low $|Z|$, seen most clearly for $s=-4$ and to a lesser extent also for $s=3$). All these features are due to selection effects, most significantly because of the completeness of the radial velocity measurements. For the nearby spatial volume, the radial velocity data is deeper in terms of absolute magnitude. The completeness gets progressively worse with greater distances, especially close to the Galactic mid-plane where the radial velocity measurements are obfuscated by stellar crowding. 

\begin{figure*}
\begin{subfigure}{.5\textwidth}
    \centering
    \includegraphics[width=.9\linewidth]{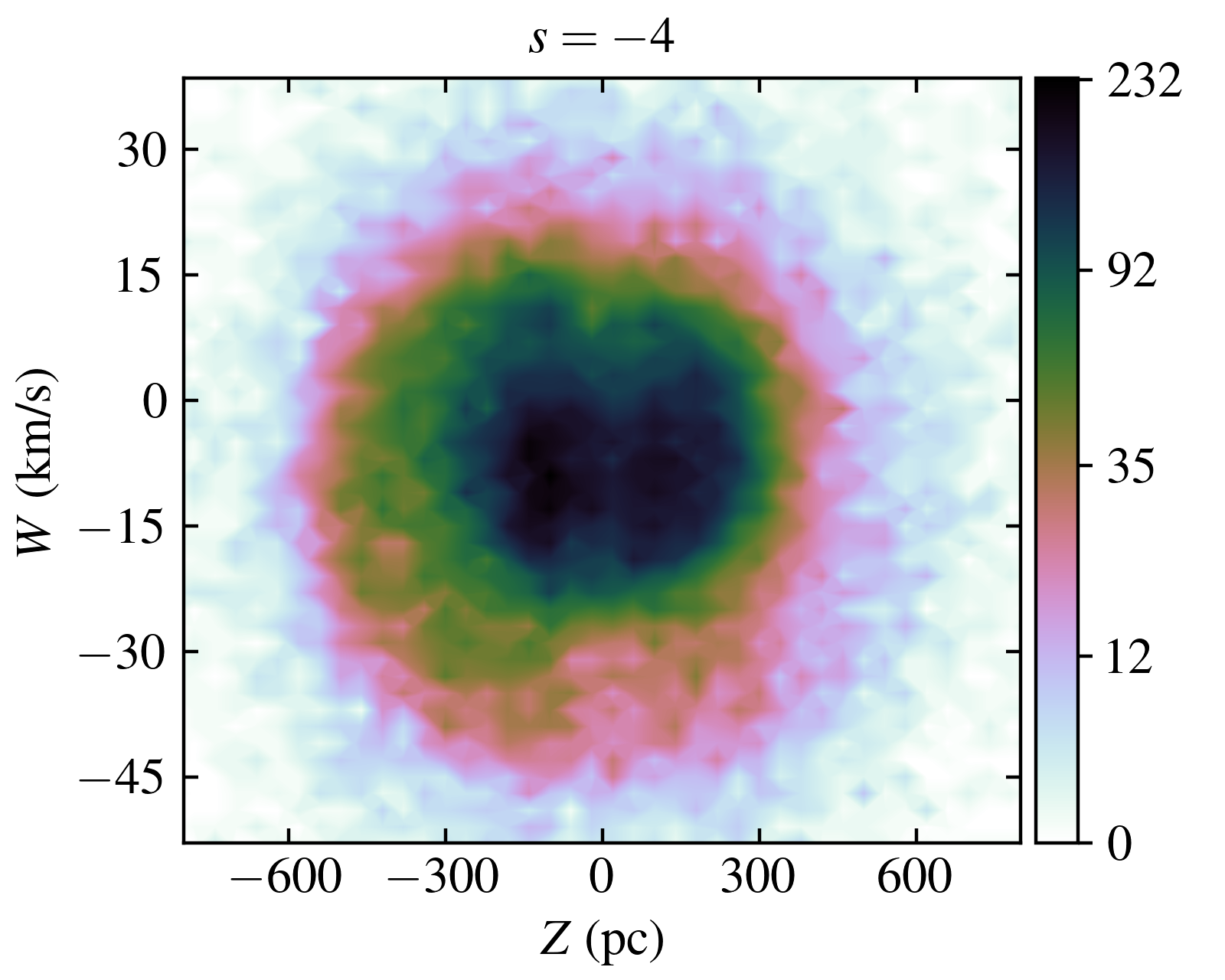}
\end{subfigure}
\begin{subfigure}{.5\textwidth}
    \centering
    \includegraphics[width=.9\linewidth]{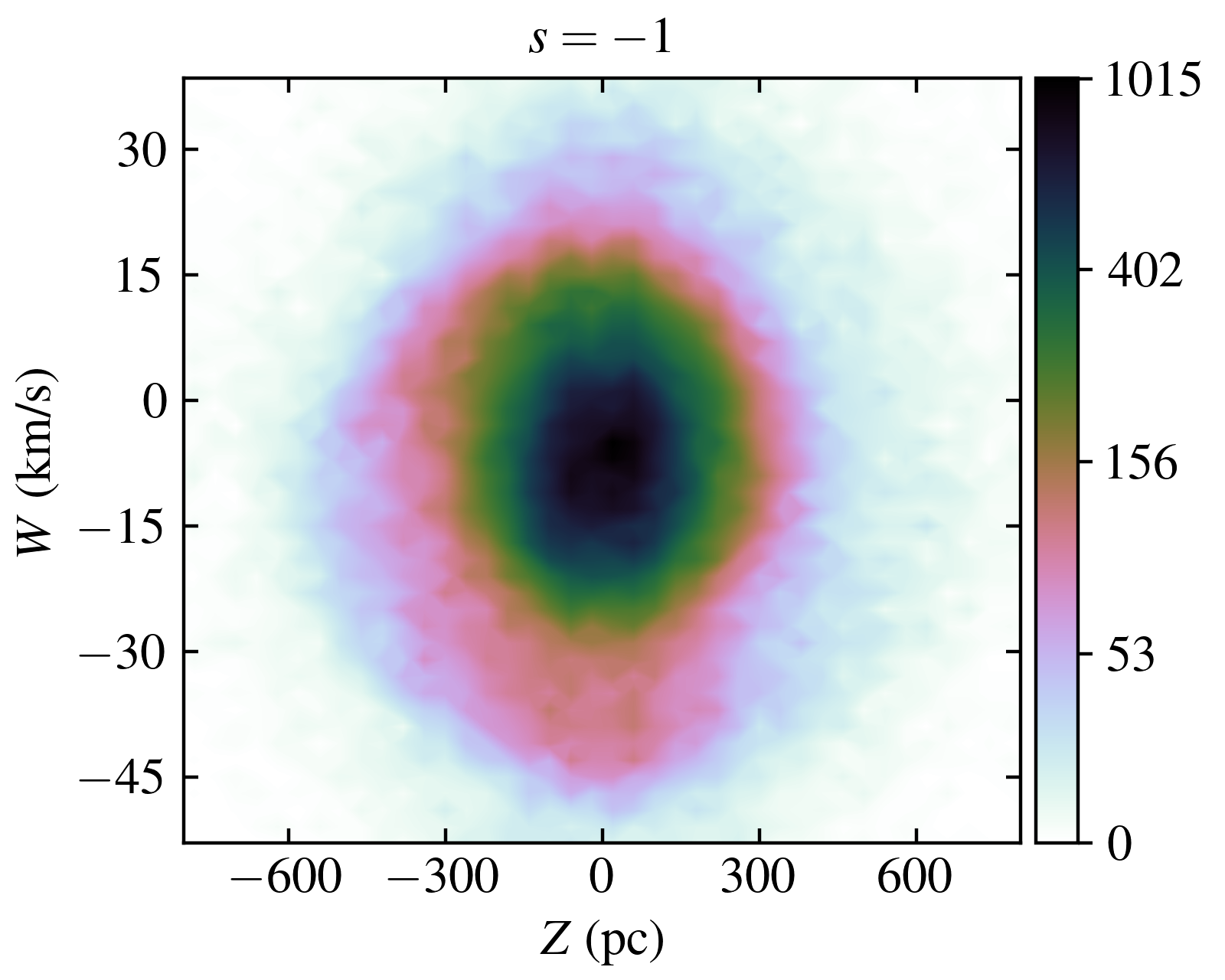}
\end{subfigure}
\par\bigskip\bigskip
\begin{subfigure}{.5\textwidth}
    \centering
    \includegraphics[width=.9\linewidth]{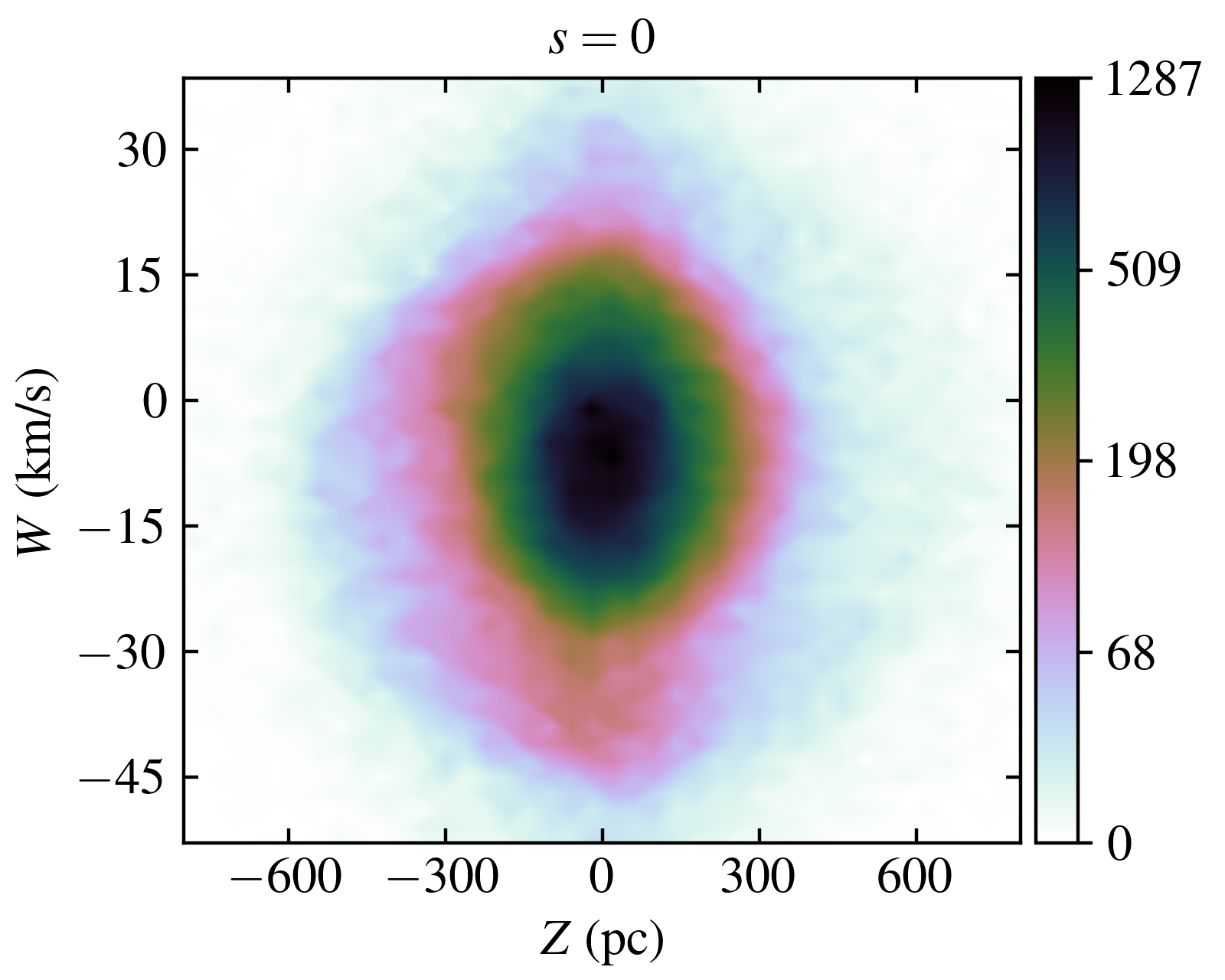}
\end{subfigure}
\begin{subfigure}{.5\textwidth}
    \centering
    \includegraphics[width=.9\linewidth]{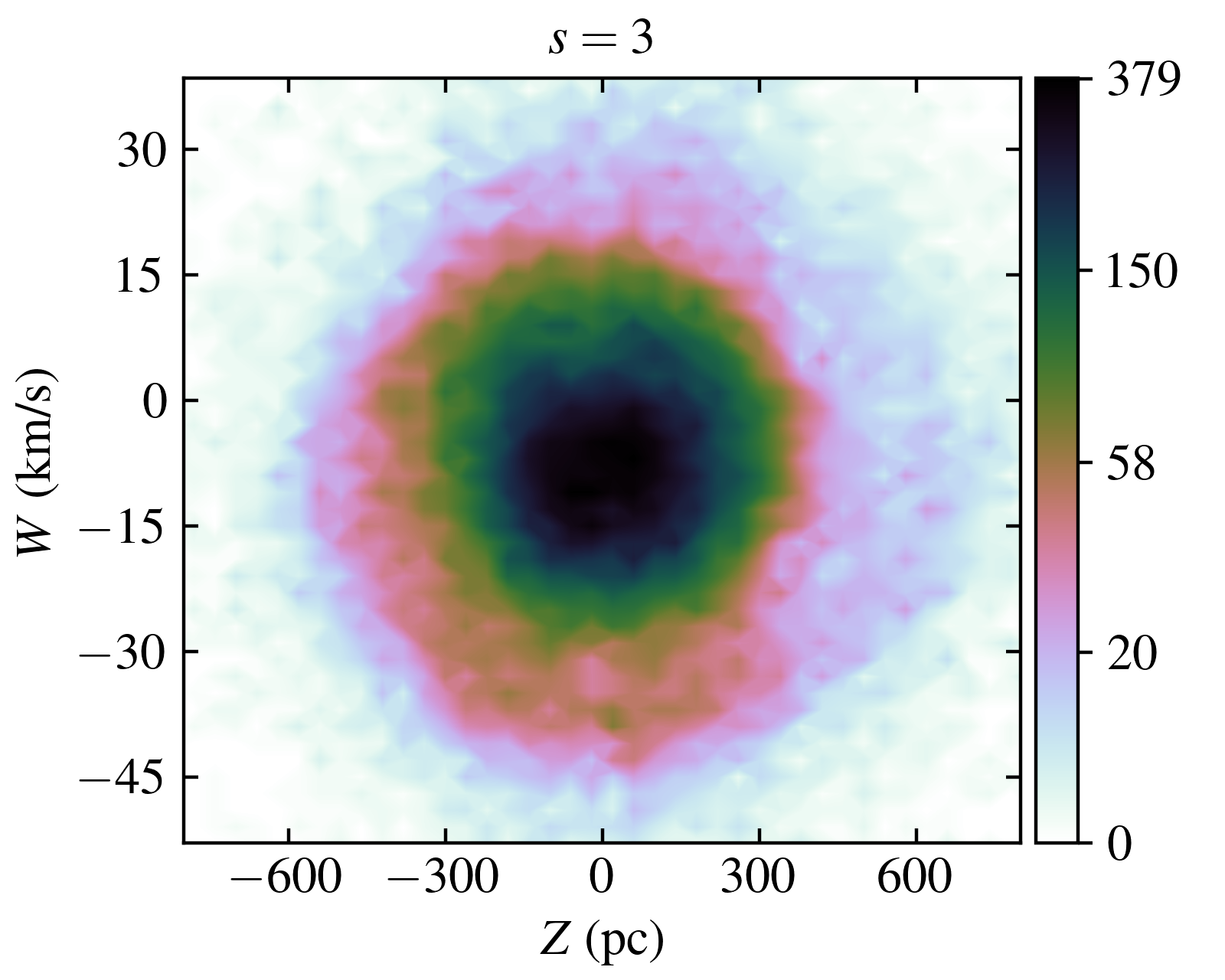}
\end{subfigure}
\caption{Histograms in the $(Z,W)$-plane, for data samples $s=\{-4,-1,0,3\}$. The scale of the colour bar is not linear, but follows $\sinh[10 \times \data_{i,j}/\text{max}(\data_{i,j})]$.}
\label{fig:hists}
\end{figure*}

\begin{figure*}
\begin{subfigure}{.5\textwidth}
    \centering
    \includegraphics[width=1.\linewidth]{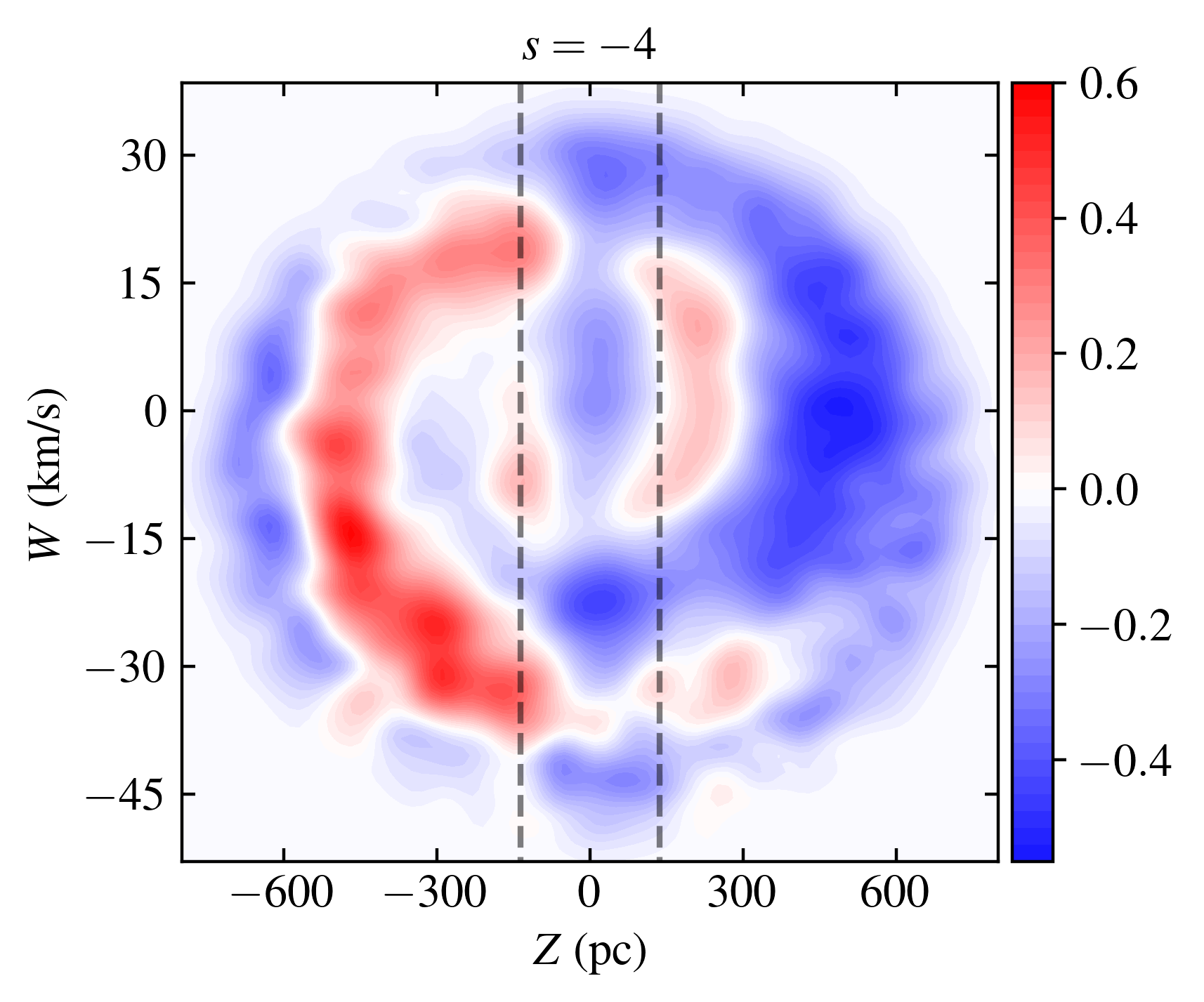}
\end{subfigure}
\begin{subfigure}{.5\textwidth}
    \centering
    \includegraphics[width=1.\linewidth]{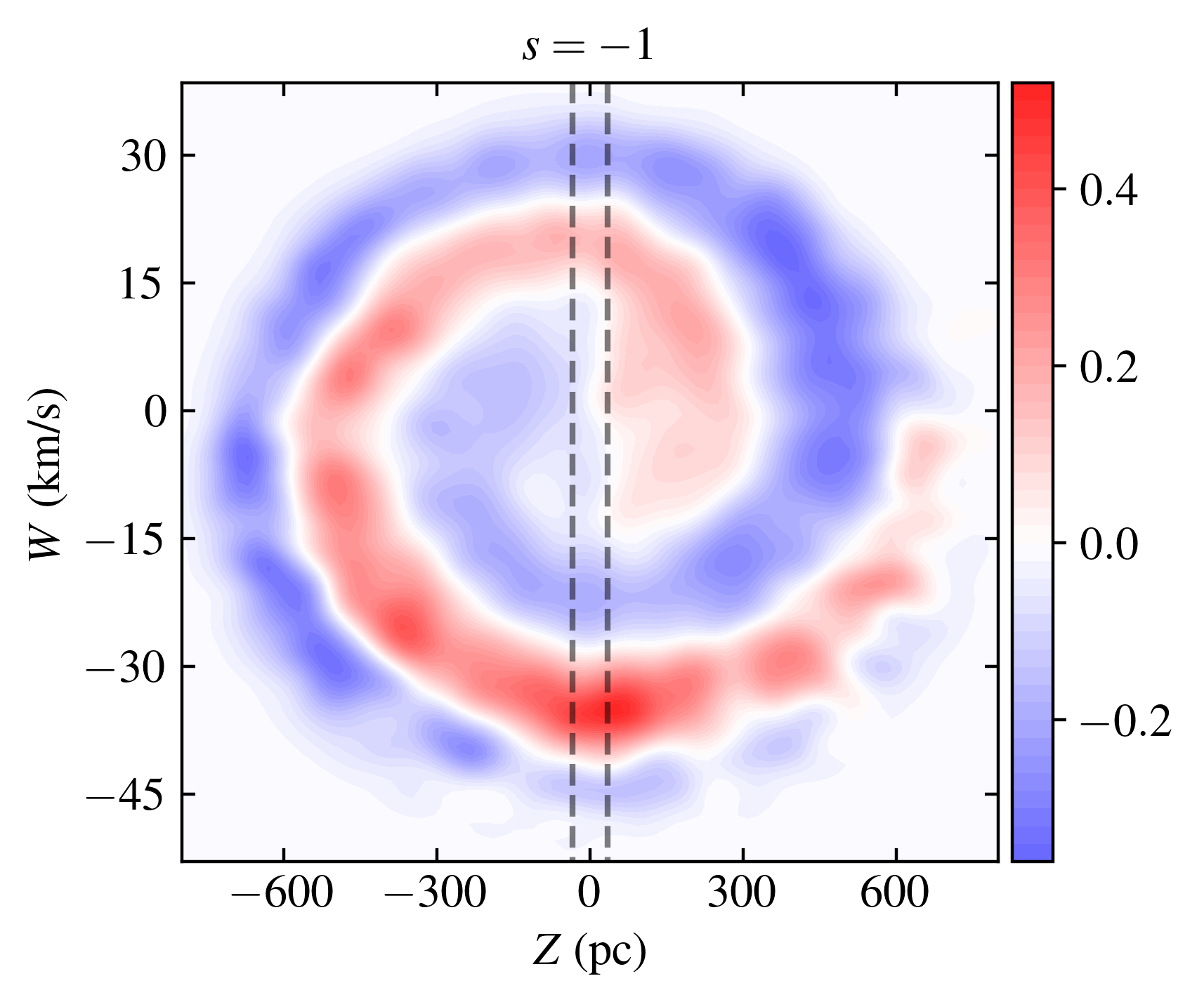}
\end{subfigure}
\par\bigskip
\begin{subfigure}{.5\textwidth}
    \centering
    \includegraphics[width=1.\linewidth]{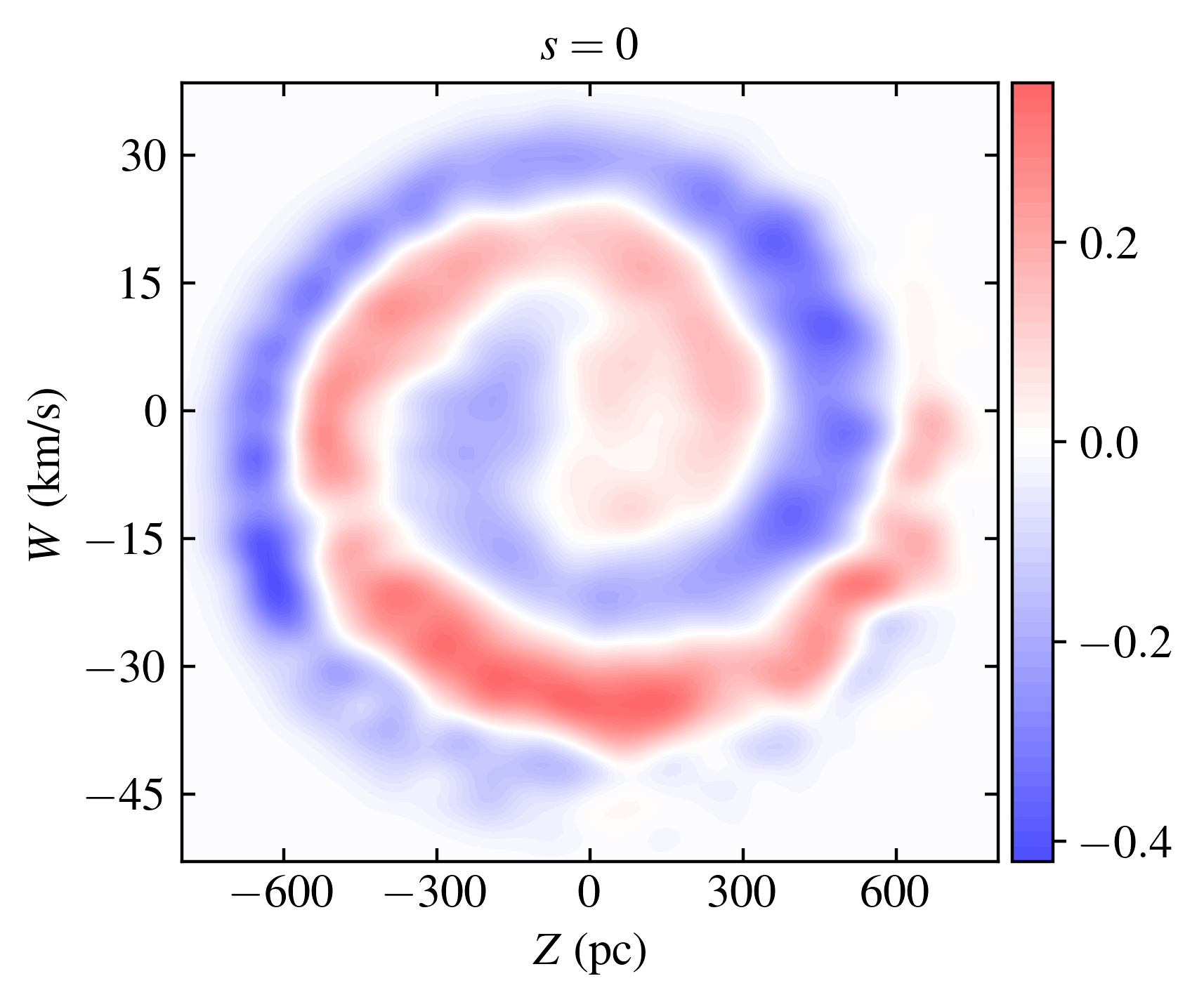}
\end{subfigure}
\begin{subfigure}{.5\textwidth}
    \centering
    \includegraphics[width=1.\linewidth]{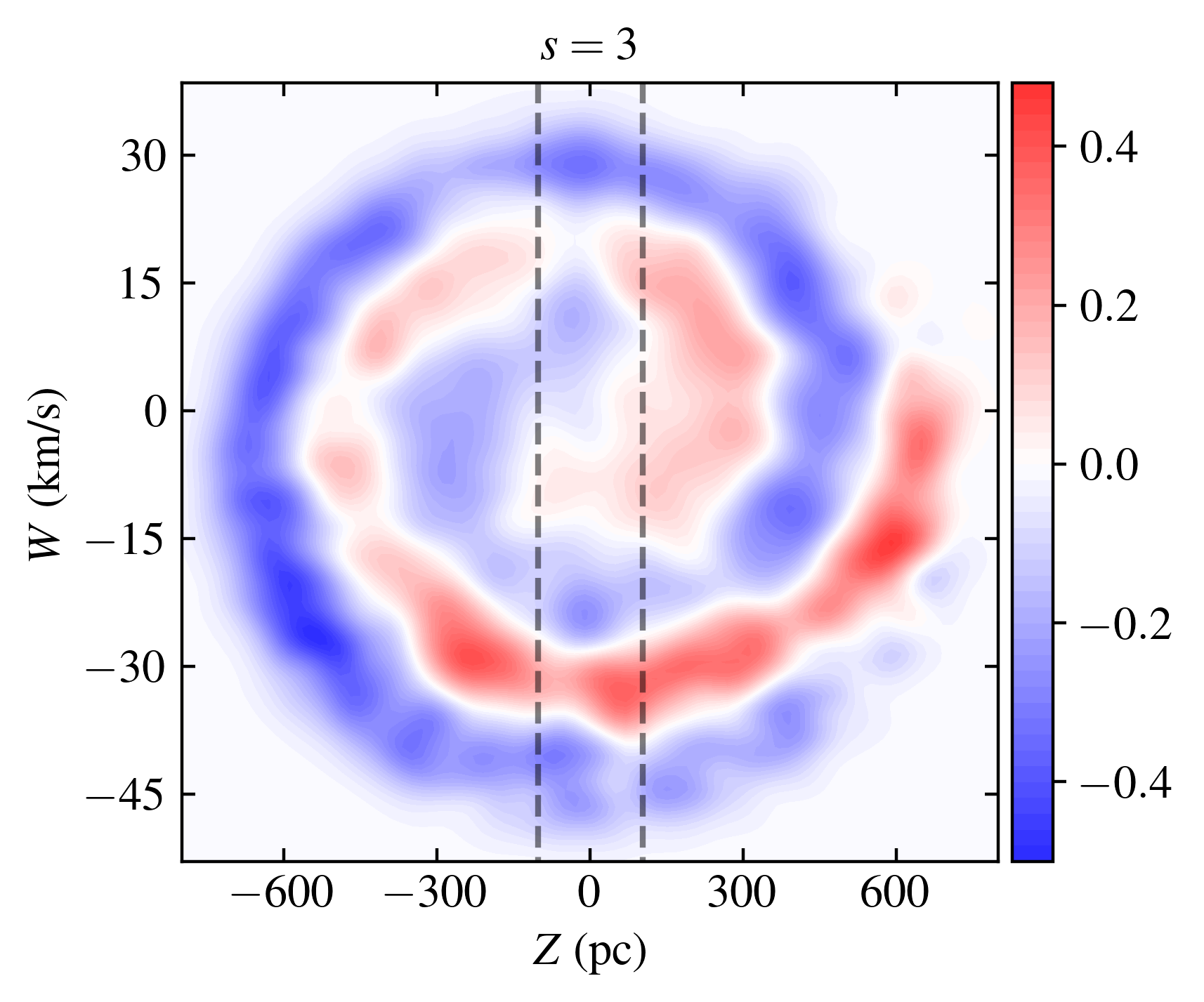}
\end{subfigure}
\caption{Spirals in the $(Z,W)$-plane, for data samples $s=\{-4,-1,0,3\}$. They are plotted in terms of the relative difference between the data histogram and the fitted bulk density distribution, as defined in Eq.~\eqref{eq:relative_dens}. The vertical dashed lines correspond to the boundaries of the $\theta$ mask function defined in Eq.~\eqref{eq:zmask}.}
\label{fig:spirals}
\end{figure*}

In Fig.~\ref{fig:spirals}, we show the phase-space spirals for the same four data samples as in Fig.~\ref{fig:hists} ($s=\{-4,-1,0,3\}$). The visualised quantity is equal to
\begin{equation}\label{eq:relative_dens}
    M(Z_i,W_j)\times\Bigg[\frac{\data_{i,j}-B(Z_i+Z_\odot,W_j+W_\odot \, | \, \popp_\text{bulk})}{B(Z_i+Z_\odot,W_j+W_\odot \, | \, \popp_\text{bulk})}\Bigg],
\end{equation}
where we have assumed a value of $Z_\odot=10~\pc$. Furthermore, this quantity is smoothed to an effective bin size of $(40~\pc)\times(2~\kmsec)$ for better visibility. The number density depression close to the Galactic mid-plane is seen clearly in this figure, and is confined to the boundaries of the $\theta$ mask function of Eq.~\eqref{eq:zmask}. There is also some asymmetry with respect to the Galactic mid-plane, at least for the more distant data samples (seen also in the figures in Appendix~\ref{app:more_plots}). It is evident from both Fig.~\ref{fig:hists} and Fig.~\ref{fig:spirals} that selection effects vary dramatically between the respective data samples. Despite this, the phase-space spirals that emerge when subtracting the bulk density are qualitatively similar, most importantly in terms of their shapes.

\begin{figure*}
\begin{subfigure}{1.\textwidth}
    \centering
    \includegraphics[width=1.\linewidth]{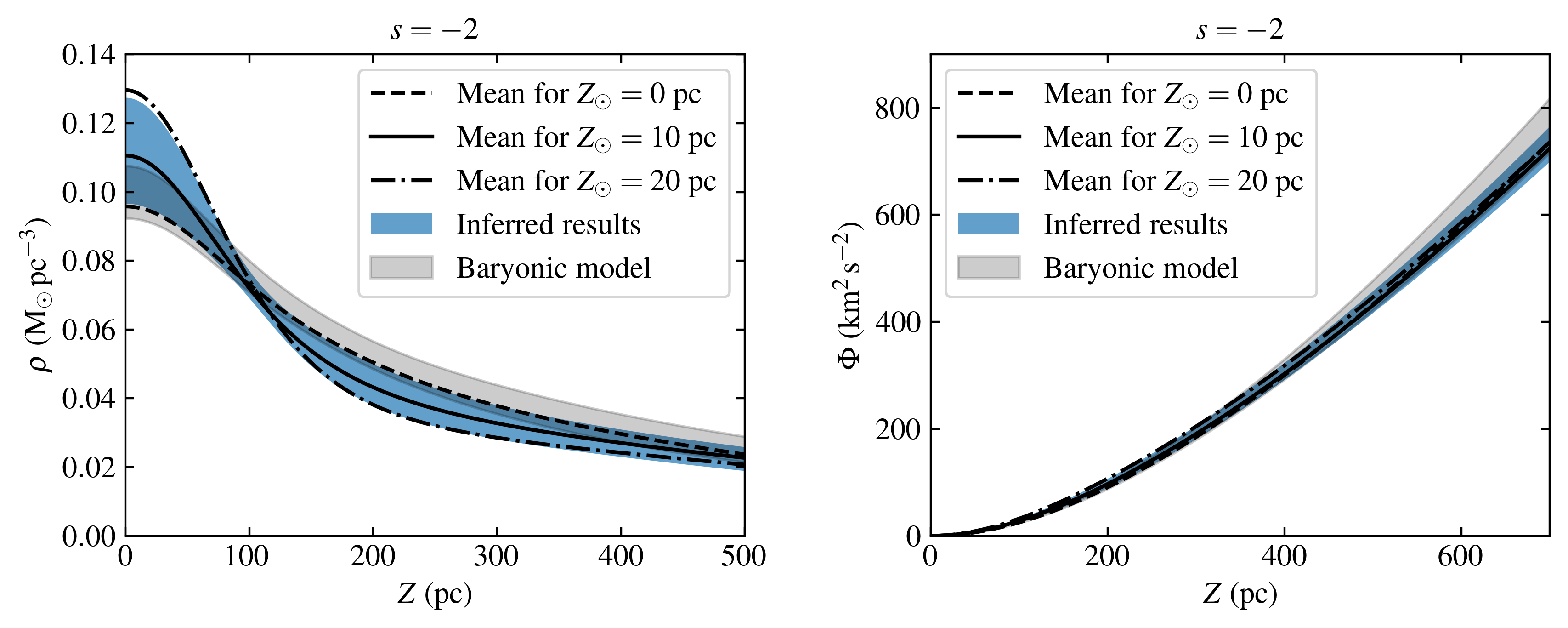}
\end{subfigure}
\par\bigskip
\begin{subfigure}{1.\textwidth}
    \centering
    \includegraphics[width=1.\linewidth]{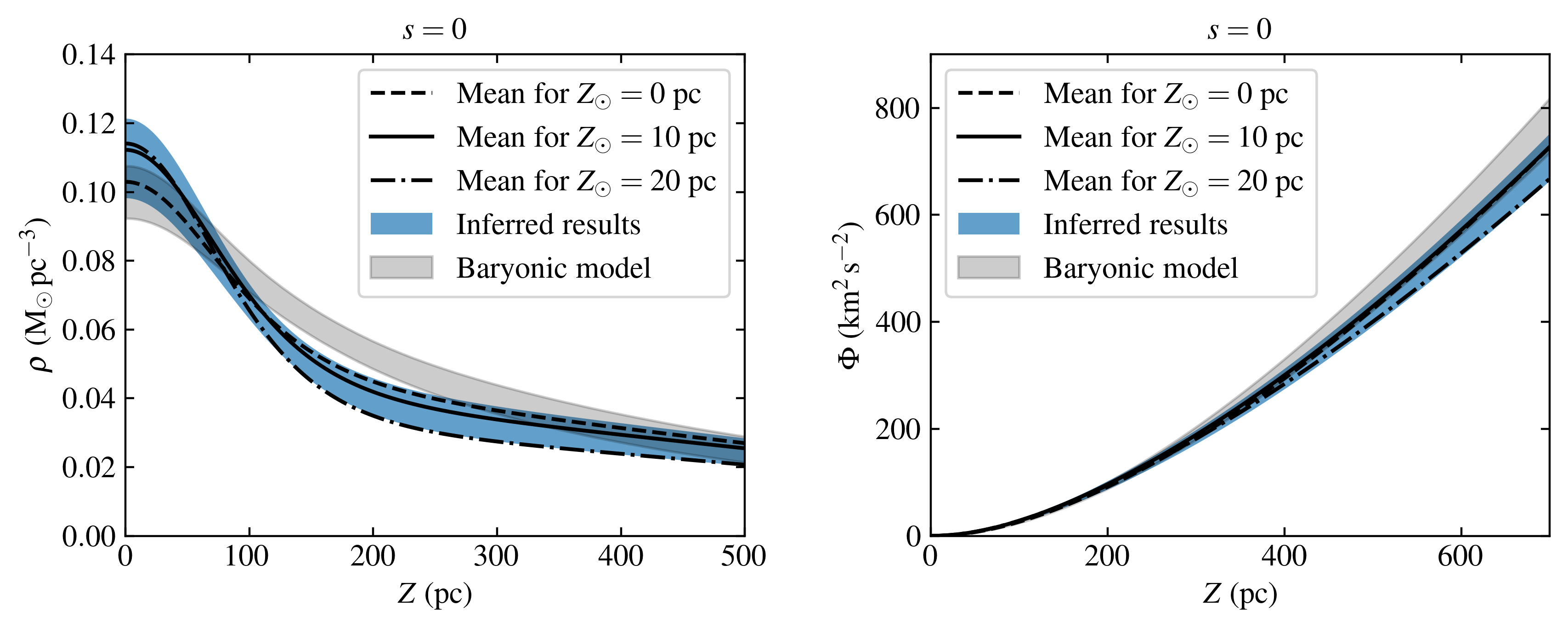}
\end{subfigure}
\par\bigskip
\begin{subfigure}{1.\textwidth}
    \centering
    \includegraphics[width=1.\linewidth]{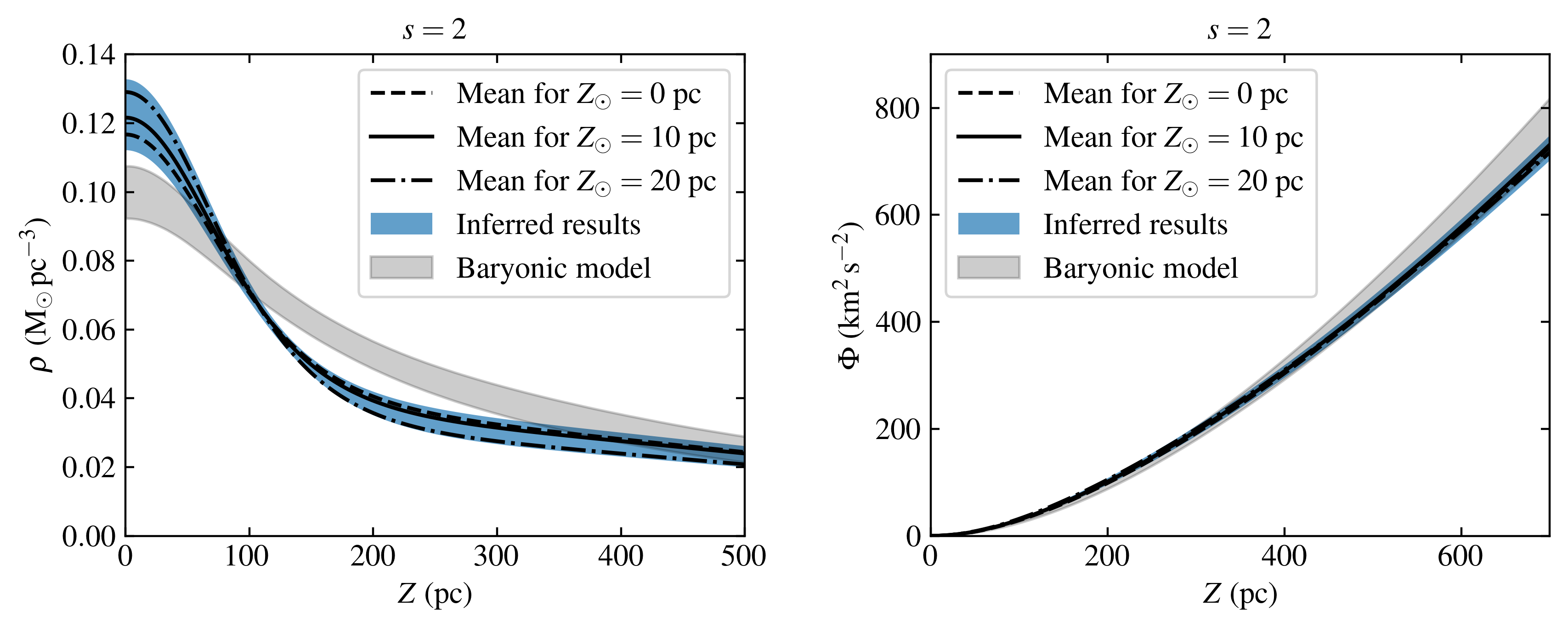}
\end{subfigure}
\caption{Inferred matter density distribution (left panels) and inferred gravitational potential (right panel), for data samples $s=\{-2,0,2\}$. The black lines correspond to the mean values for different fixed values of $Z_\odot$. We also show $1\sigma$ bands of the inferred results and the baryonic density distribution model.}
\label{fig:jks}
\end{figure*}

In Fig.~\ref{fig:jks}, we show the inferred matter density distribution and inferred gravitational potential for three representative data samples ($s=\{-2,0,2\}$). The coloured band corresponds to the $1\sigma$ statistical uncertainty given by jackknifing (as explained in Sect.~\ref{sec:jackknife}), including the variance coming from the three different solar height values. The black lines correspond to the mean result for the three different solar heights ($Z_\odot = \{0,10,20\}~\pc$). The grey band shows the baryonic model to within a $1\sigma$ uncertainty, including a local dark matter density of $0.011\pm0.003~\Msunppcc$. The inferred gravitational potential agrees well with the baryonic model, and is very well consistent between samples.

The inferred matter density distribution has a higher relative statistical uncertainty than the inferred gravitational potential, which is expected in any dynamical mass measurement due to it being derived from the gravitational potential's second order derivative. It also varies more dramatically between samples, and also for the different values of $Z_\odot$. A general trend seen in Fig.~\ref{fig:jks}, as well as for the other data samples shown in Appendix~\ref{app:more_plots}, is that the inferred matter density distribution is flatter and agrees better with the baryonic model for the data samples closer to the Galactic centre ($s\leq1$). Conversely, the data samples in the direction of the anti-centre ($s\geq2$) are more pinched, and somewhat discrepant with the baryonic model. There is a similar but less pronounced trend for the different values of $Z_\odot$, where especially $Z_\odot=0~\pc$ gives rise to flatter matter density distributions.

\begin{figure*}
	\includegraphics[width=.95\textwidth]{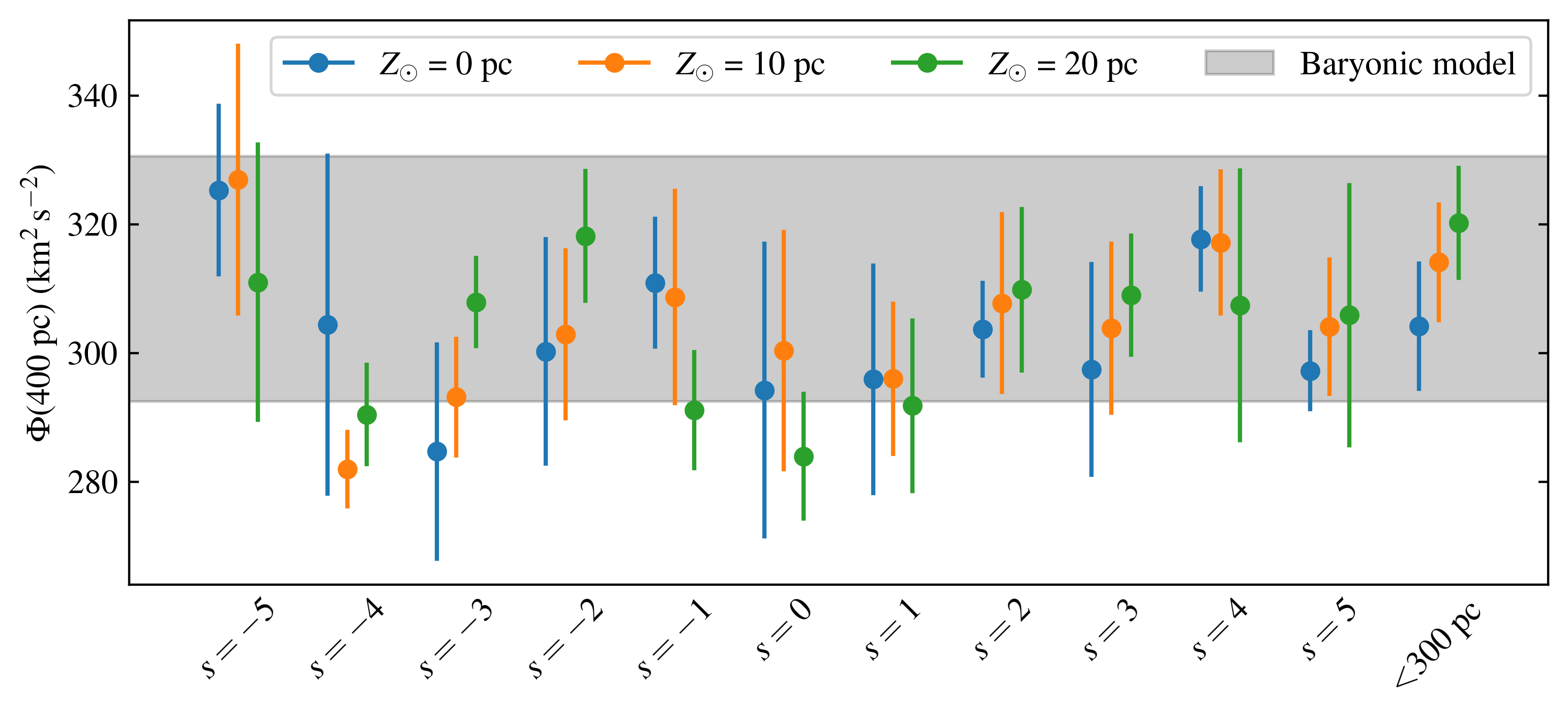}
    \caption{Inferred gravitational potential at height $z=400~\pc$ for all data samples ($s$ from -5 to 5 and $\sqrt{X^2+Y^2}<300~\pc$). The markers and vertical lines show the mean and standard deviations as derived from jackknifing, where the colours represent the three fixed values for the height of the Sun ($Z_\odot = \{0,10,20\}~\pc$). The grey band represents the baryonic model, including a dark matter density of $0.011\pm0.003 ~\Msunppcc$.}
    \label{fig:summary_phi400pc}
\end{figure*}

In Fig.~\ref{fig:summary_phi400pc}, we show the inferred value of $\Phi(400~\pc)$ for all twelve data samples. As suggested by the test on one-dimensional simulations in \citetalias{PaperI}, the gravitational potential at this height is the quantity that is most robustly inferred. Indeed, the inferred values for $\Phi(400~\pc)$ are largely consistent between the different data samples. Interestingly, this is the case even for the samples at greater distances (i.e. large $|s|$), for which the spiral was much less clearly defined (e.g. $s=-4$ in Fig.~\ref{fig:spirals}). This demonstrates that our method works reasonably well even when the data is noisy and suffers from severe selection effects, which are absorbed by the bulk density distribution. It is expected that the mass of the Galactic disk varies with Galactocentric radius, giving rise to a negative slope with respect to the data sample index $s$. Such a slope is allowed but not preferred given our results, and varies depending on what data samples are considered reliable enough to include in the fit of such a slope. For some data samples, the size of the statistical uncertainties differ quite a lot depending on $Z_\odot$. This is most notable for data sample $s=-4$, which has a very large uncertainty for $Z_\odot = 0~\pc$, but rather small uncertainties for $Z_\odot = \{10,20\}~\pc$. The reason is that the inner mask function, $m(z,w\,|\,\rho_h)$ as defined in Eq.~\eqref{eq:inner_boundary}, is a function of the phase-space coordinates in the rest frame of the Galactic disk. This means that this masked region shifts somewhat in the $(Z,W)$ histogram, depending on the fixed value of $Z_\odot$. Especially for the noisy histogram of data sample $s=-4$, this small shift has a significant effect on the likelihood function.

In order to infer the local dark matter density, we compared the inferred value of $\Phi(400~\pc)$ with that of the baryonic model. The baryons and the matter density correction of Eq.~\eqref{eq:rho_correction} amounted to a total contribution of $265.0 \pm 16.0~\kmsecsq$ to $\Phi(400~\pc)$. Any surplus with respect to this value was interpreted as halo dark matter. The reason for only using the value of $\Phi(400~\pc)$ is that this quantity was shown to be the most accurately inferred in tests on one-dimensional simulations \citepalias{PaperI}, and also agrees well between the different data samples and fixed values of $Z_\odot$ used in this work.

When placing constraints on the surface density of a thin dark disk, we used a similar line of reasoning: a thin dark disk was allowed to the extent that there is room for its contribution to $\Phi(400~\pc)$. We use the same model for the baryonic matter densities, and a dark matter halo density coming from independent studies of the Milky Way circular velocity curve (to use the halo density derived in this or any other work that analyses the vertical motion of stars would be circular reasoning). In order to be conservative when placing thin dark disk constraints, we assumed a local halo dark matter density of $\rho_{\text{DM},\odot} = 0.009 \pm 0.003~\Msunppcc$, which is a somewhat low value with a large uncertainty.\footnote{In a recent review on the local dark matter density by \cite{2020arXiv201211477D}, the measurements based on the Milky Way circular velocity curve lie in the range $0.008$--$0.013~\Msunppcc$, which is lower than the range of $0.011$--$0.016~\Msunppcc$ for more local analyses. For example, \cite{2019JCAP...10..037D} report $0.008 \pm 0.0008~\Msunppcc$ or $0.001 \pm 0.0001~\Msunppcc$ depending on the baryonic model, \cite{2020MNRAS.494.4291C} report $0.009 \pm 0.0005~\Msunppcc$, and \cite{2020PhRvD.102l3028P} reports $0.0096 \pm 0.0005~\Msunppcc$ or $0.0010 \pm 0.0005~\Msunppcc$ depending on the shape of the dark matter halo. With this in mind, $\rho_{\text{DM},\odot} = 0.009 \pm 0.003~\Msunppcc$ is a conservative choice when constraining the thin dark disk surface density, because the assumed $\rho_{\text{DM},\odot}$ has a low mean and a large uncertainty.}
A thin dark disk with a surface density of $\Sigma_\text{DD} = 1~\Msunppcsquare$ and a scale height of $h_\text{DD}=\{20,\,50,\,100\}~\pc$, assuming a matter density profile
\begin{equation}
    \rho_\text{DD} = \frac{\Sigma_\text{DD}}{4h_\text{DD}}
    \cosh^{-2}\Bigg(\frac{z}{2h_\text{DD}}\Bigg),
\end{equation}
contributes with $\{9.9,\,8.9,7.2\}~\kmsecsq$ to $\Phi(400~\pc)$.

When inferring the local dark matter density, as well as setting constraints on the thin dark disk surface density, we only use the information coming from the nearby data samples, namely $|s| \leq 3$. The reason for this is that the more distant data samples have less statistics and suffer from more extreme selection effects. This is seen in for example Fig.~\ref{fig:spirals}, where the spiral of data sample $s=-4$ is poorly defined. Our method does seem to have worked well even for those data samples, but we still found it more cautious to exclude them. When summarising the results of several data samples and a fixed value of $Z_\odot$, the total weighted mean was calculated according to
\begin{equation}
    \{\text{weighted mean}\} = \frac{ \sum_{s=-3}^{3} \{\text{mean}\}_s \times \{\text{variance}\}_s^{-1}}{\sum_{s=-3}^{3}\{\text{variance}\}_s^{-1}},
\end{equation}
with a total statistical variance of
\begin{equation}
    \{\text{total variance}\} = \frac{1}{\sum_{s=-3}^{3}\{\text{variance}\}_s^{-1}}.
\end{equation}
For the respective values of $Z_\odot$, the summary results of data samples $|s|\leq3$ was
\begin{equation}
	\Phi(400~\pc) = 
	\begin{cases}
		304.8 \pm 4.0~\kmsecsq, \quad \text{for} \, Z_\odot=0~\pc, \\
		295.3 \pm 3.7~\kmsecsq, \quad \text{for} \, Z_\odot=10~\pc, \\
		305.2 \pm 3.9~\kmsecsq, \quad \text{for} \, Z_\odot=20~\pc.
	\end{cases}
\end{equation}
These three different values of $Z_\odot$ have a total mean value of $301.8~\kmsecsq$, and the dispersion between them amounts to $4.6~\kmsecsq$. In order to account for the statistical variance given a fixed value of $Z_\odot$, as well as the variance that arises from changing $Z_\odot$, we added these measurement uncertainties together according to
\begin{equation}
	\sqrt{\frac{4.0^2+3.7^2+3.9^2}{3}+4.6^2}\,\kmsecsq = 6.0~\kmsecsq.
\end{equation}
Hence, for the joint analysis of samples $|s| \leq 3$, we inferred $\Phi(400~\pc) = 301.8 \pm 6.0~\kmsecsq$ and $\rho_{\text{DM},\odot} = 0.0085 \pm 0.0039~\Msunppcc$. The dominant component of this uncertainty comes from the baryonic model ($0.0037~\Msunppcc$), which was added in quadrature to the statistical measurement uncertainty ($0.0014~\Msunppcc$).

Using the assumptions discussed above and a dark disk scale height of $50~\pc$, the likelihood of its surface density is a Gaussian with a mean and standard deviation of $-0.24 \pm 2.40~\Msunppcsquare$, which is consistent with zero. Negative values for this quantity are unphysical and therefore excluded; the upper limits are given by the likelihood ratio relative to the null hypothesis (i.e. $\Sigma_\text{DD} = 0~\Msunppcsquare$).
We obtain an upper $68~\%$ ($95~\%$) confidence limit of $2.17~\Msunppcsquare$ ($4.56~\Msunppcsquare$). For scale heights of $20~\pc$ and $100~\pc$, the limits are $1.96~\Msunppcsquare$ ($4.13~\Msunppcsquare$) and $2.71~\Msunppcsquare$ ($5.59~\Msunppcsquare$), respectively. Even if we would have been extremely conservative and assumed no halo dark matter, we would still have placed the most stringent constraints on a thin dark disk surface density, corresponding to an upper $68~\%$ ($95~\%$) confidence limit of $6.02~\Msunppcsquare$ ($7.93~\Msunppcsquare$), assuming a scale height of $50~\pc$.

As a comparison, for the data sample with a spatial cut ($\sqrt{X^2+Y^2}<300~\pc$), the inferred values were
\begin{equation}
	\Phi(400~\pc) = 
	\begin{cases}
		304.2 \pm 9.7~\kmsecsq, \quad \text{for} \, Z_\odot=0~\pc, \\
		314.1 \pm 9.0~\kmsecsq, \quad \text{for} \, Z_\odot=10~\pc, \\
		320.3 \pm 8.5~\kmsecsq, \quad \text{for} \, Z_\odot=20~\pc.
	\end{cases}
\end{equation}
Using the same analysis as above, applied to this one data sample, we obtained $\Phi(400~\pc) = 312.9 \pm 11.2$ and $\rho_{\text{DM},\odot} = 0.0111 \pm 0.0045~\Msunppcc$. This result is statistically consistent with the one presented above, although slightly higher in both value and uncertainty. For a thin dark disk with a scale height of $50~\pc$, the upper $68~\%$ ($95~\%$) confidence limit is $3.62~\Msunppcsquare$ ($6.25~\Msunppcsquare$).

\section{Discussion}\label{sec:discussion}

Using a new method that extracts information from the time-varying structure of the Milky Way phase-space spiral, we have been able to infer the vertical gravitational potential of the solar neighbourhood for eleven statistically independent data samples. Using these results and a model for the total matter density of baryons, we have inferred the local halo dark matter density, as well as placed the most stringent constraints on the surface density of a thin dark disk.

Overall, the inferred gravitational potential of the respective data samples agree well, at least in terms of $\Phi(400~\pc)$, which was deemed to be the most robustly inferred quantity in tests on one-dimensional simulations in \citetalias{PaperI}. The results are consistent, despite the fact that the data samples are subject to very different selection effects and differ significantly in their number density profiles as a function of height. For data sample $s=0$, we see a high stellar number density, especially close to the Galactic mid-plane; conversely, for the distant data samples, most notably $|s| \geq 4$, the mid-plane number density is low. Despite this, the different data samples gave rise to very similar results; this illustrates that our method is robust with respect to such systematics, and that selection effects are absorbed by the bulk density distribution.

The inferred matter density distributions agree fairly well with the baryonic model, especially for $Z_\odot=0~\pc$ and $s\leq1$. This can be compared with \cite{Widmark2019} and \cite{2021A&A...646A..67W}, where they used a similarly free model for the total density, but weighed the Galactic disk using a distribution function fitting method based on the assumption of a steady state. For several statistically independent stellar samples, they inferred a matter density distribution that was very pinched (i.e. with a high mid-plane value and quickly decreasing with height) and argued that their result must be biased by time-varying dynamics. The reason that the results of this work agree better with the baryonic model is probably explained by the fact that the different methods are subject to different systematic biases. For example, the method used in this work is not as sensitive to the distribution of stars with low vertical energies ($E_z \lesssim \Phi(300~\pc)$).

Using summary statistics of seven data samples (fulfilling $|s|\leq3$), we infer a local halo dark matter density of $\rho_{\text{DM},\odot} = 0.0085 \pm 0.0039~\Msunppcc$. This is a somewhat low value compared to most other analyses of the vertical motion of stars in the solar neighbourhood (although strictly speaking not discrepant as our uncertainty is fairly large), and more in line with more global estimates, such as those coming from the Milky Way rotational velocity curve \citep{2020arXiv201211477D}. We also set an upper $68~\%$ ($95~\%$) confidence limit of $2.17~\Msunppcsquare$ ($4.56~\Msunppcsquare$) to the matter density of a thin dark disk with a scale height $\leq 50~\pc$. This is significantly stronger than any previous constraints on a thin dark disk (\citealt{Schutz:2017tfp} set an upper $95~\%$ confidence limit of roughly $10~\Msunppcsquare$, for a scale height of $50~\pc$). The statistical uncertainties on the inferred halo dark matter density and thin dark disk surface density are dominated by the uncertainty associated with the baryonic model. Therefore, in order to make further progress, it is crucial to revisit and improve on the baryonic matter density distributions of the solar neighbourhood. As discussed in Sect.~\ref{sec:baryonic_model}, it is not implausible that the baryonic model used in this work suffers from significant systematic errors, potentially larger than the reported statistical uncertainties.

For the data sample that only had a cut in spatial coordinates, according to $\sqrt{X^2+Y^2}<300~\pc$, the results agree with the main analysis to within $1\sigma$, both in terms of $\Phi(400~\pc)$ and $\rho_{\text{DM},\odot}$ (where only the latter includes uncertainties from the baryonic model). The results agree particularly well for $Z_\odot = 0~\pc$. This data sample is not statistically independent from the main analysis and was only included as a test of consistency. We consider this result less reliable, due to not having any cuts in angular momentum and thus including stars with strongly elliptical orbits.

In this work, we did not attempt to infer the height of the Sun with respect to the Galactic mid-plane ($Z_\odot$), mainly because of strong selection effects. Rather, we accounted for the uncertainty of this parameter by producing results for three different fixed values of $Z_\odot = \{0,10,20\}~\pc$. Interestingly, studies based on the more local spatial volume, within a few hundred parsec, tend to prefer lower values ($0$--$10~\pc$, e.g. \citealt{Buch:2018qdr,2021A&A...646A..67W,2021A&A...649A...6G}); conversely, higher values typically come from studies that reach several kilo-parsec in height (e.g. \citealt{Juric:2005zr,2017MNRAS.468.3289Y,BovyAssym}). This discrepancy indicates that the Galaxy is not perfectly mirror symmetric and that an estimate of $Z_\odot$ can depend on how this quantity is defined.
With this caveat in mind, a comparison between the baryonic model and the inferred matter density distributions seems to suggest that our results are vaguely in favour of $Z_\odot=0~\pc$, which is in agreement with other local studies.

The spatial reach of the method employed in this work is mainly limited by the completeness of the radial velocity sample. This will improve with \emph{Gaia}'s future data releases, most immanently with the full third data release (EDR3 only contains a cleaned version of the second data release's radial velocity sample). Apart from such improvements to the data, it should be possible to use the astrometric information of stars even when the radial velocity is missing; especially for more distant parts of the Galactic disk, stars close to the Galactic mid-plane have a vertical velocity that is well approximated by their latitudinal proper motion. With careful data treatment, we might be able to circumvent the issue of missing radial velocity measurements and weigh different parts of the Galactic disk with high precision.

In order to perform more careful tests and further refine this new method, we plan to apply it to high resolution three-dimensional galaxy simulations. To reproduce well-resolved phase-space spirals requires simulations with roughly a billion particles, which only very recently have become feasible (\citealt{2020MNRAS.499.2416A}, Hunt et al. in preparation).

\section{Conclusion}\label{sec:conclusion}

For the first time, we have employed a method for weighing the Galactic disk using the time-varying structure of the Milky Way phase-space spiral. Our method extracts information from the shape of the phase-space spiral and is complementary to traditional methods that are based on the assumption of a steady state. Using a baryonic model, we have inferred a local halo dark matter density of $\rho_{\text{DM},\odot} = 0.0085 \pm 0.0039~\Msunppcc = 0.32 \pm 0.15~\GeVcmcc$, which is consistent with other recent measurements. Using conservative assumptions, we have been able to place the most stringent constraints on the surface density of a thin dark disk: a $95~\%$ confidence limit of roughly $5~\Msunppcsquare$, assuming a dark disk scale height $\leq 50~\pc$.

For both the halo dark matter density and the surface density of a thin dark disk, the statistical uncertainty is dominated by the baryonic model. As discussed in Sect.~\ref{sec:baryonic_model}, this model is somewhat outdated and could potentially suffer from significant systematic errors, both in terms of its stellar and gaseous components. We plan to update this model using new data in the near future.

Our method places strong constraints on the weight of the Galactic disk and its dark sector components. In terms of its precision, it is highly competitive with respect to methods based on the assumption of a steady state. In a general sense, this illustrates that time-varying structures, that break the assumption of a steady state, are not solely obstacles to dynamical mass measurements, but can in fact be assets containing useful information.

\begin{acknowledgements}
We thank the anonymous referee for a constructive review. AW acknowledges support from the Carlsberg Foundation via a Semper Ardens grant (CF15-0384).
CL acknowledges funding from the European Research Council (ERC) under the European Union’s Horizon 2020 research and innovation programme (grant agreement
No. 852839).
PFdS acknowledges support by the Vetenskapsr{\aa}det (Swedish Research Council) through contract No. 638-2013-8993 and the Oskar Klein Centre for Cosmoparticle Physics.
GM acknowledge funding from the Agence Nationale de la Recherche (ANR project ANR-18-CE31-0006 and ANR-19-CE31-0017) and from the European Research Council (ERC) under the European Union’s Horizon 2020 research and innovation programme (grant agreement No. 834148).
This work made use of an HPC facility funded by a grant from VILLUM FONDEN (projectnumber 16599).
This work was supported in part by World Premier International Research Center Initiative (WPI Initiative), MEXT, Japan.

This work has made use of data from the European Space Agency (ESA) mission \emph{Gaia} (\url{https://www.cosmos.esa.int/gaia}), processed by the \emph{Gaia} Data Processing and Analysis Consortium (DPAC,
\url{https://www.cosmos.esa.int/web/gaia/dpac/consortium}). Funding for the DPAC has been provided by national institutions, in particular the institutions participating in the \emph{Gaia} Multilateral Agreement.

This research utilised the following open-source Python packages: \textsc{Matplotlib} \citep{matplotlib}, \textsc{NumPy} \citep{numpy}, \textsc{SciPy} \citep{scipy}, \textsc{Pandas} \citep{pandas}, \textsc{TensorFlow} \citep{tensorflow2015-whitepaper}.
\end{acknowledgements}




\nocite{PaperOne}
\bibliographystyle{aa} 
\bibliography{thisbib} 

\begin{appendix} 

\section{Plots of the remaining data samples}\label{app:more_plots}

In this appendix, we present the plots corresponding to Figs.~\ref{fig:hists}, \ref{fig:spirals}, and \ref{fig:jks}, for the data samples that were not already shown in Sect.~\ref{sec:results}. We show the data histograms in Fig.~\ref{fig:hists_app1}, the extracted spirals in Fig.~\ref{fig:spirals_app1}, and the inferred matter density distribution and gravitational potential in Fig.~\ref{fig:jks_app1}.

\begin{figure*}
\begin{subfigure}{.5\textwidth}
    \centering
    \includegraphics[width=.95\linewidth]{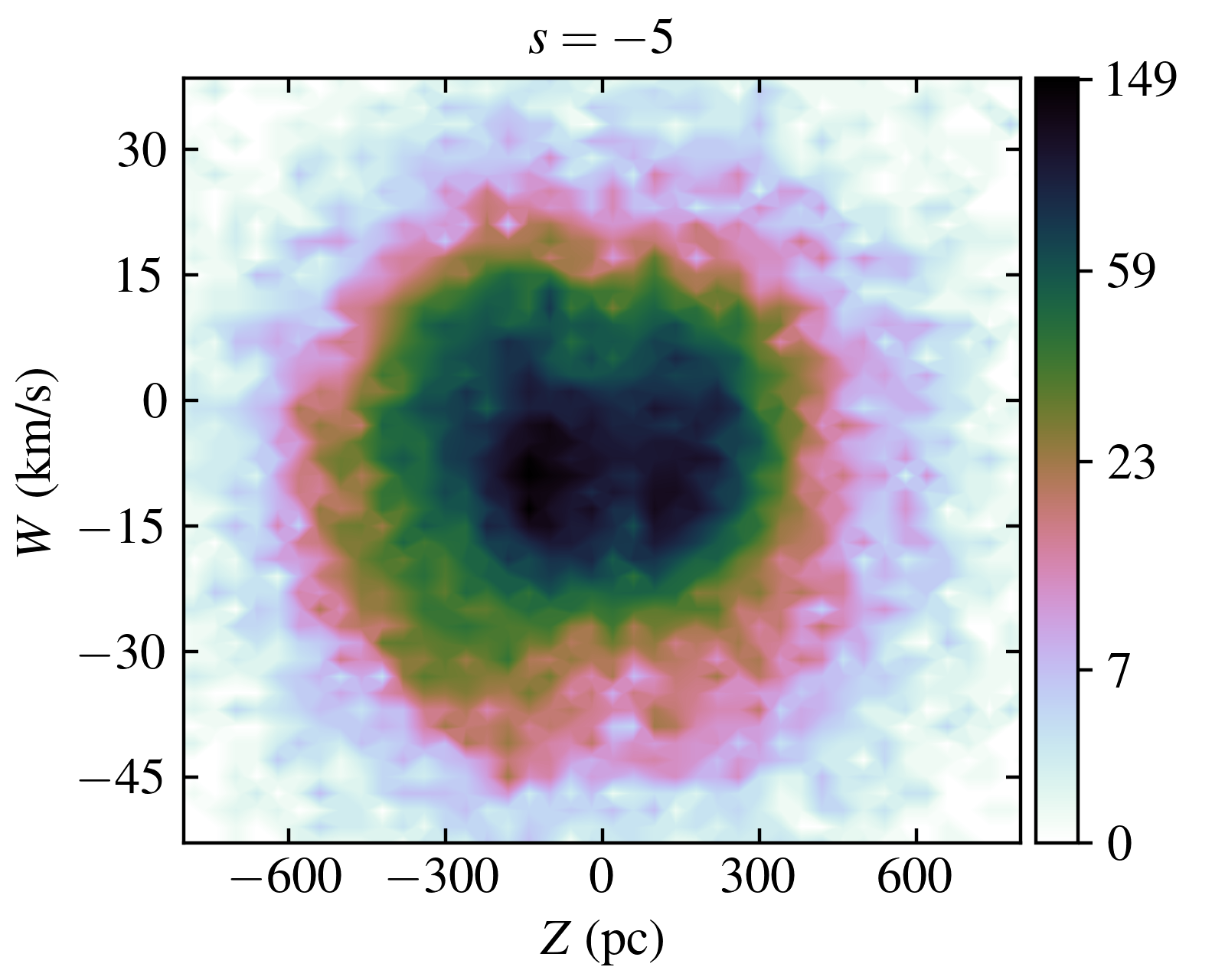}
\end{subfigure}
\begin{subfigure}{.5\textwidth}
    \centering
    \includegraphics[width=.95\linewidth]{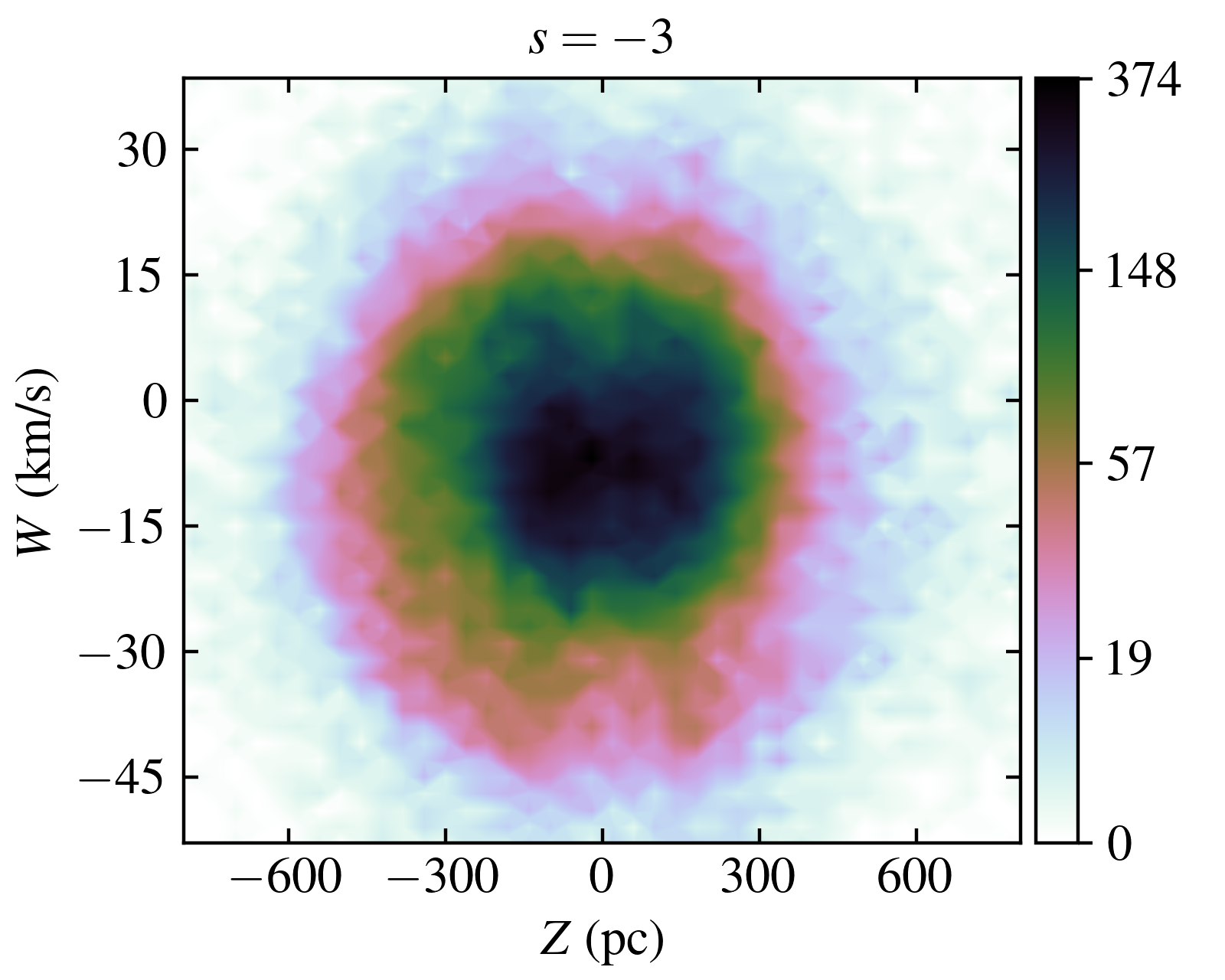}
\end{subfigure}
\par\bigskip\bigskip
\begin{subfigure}{.5\textwidth}
    \centering
    \includegraphics[width=.95\linewidth]{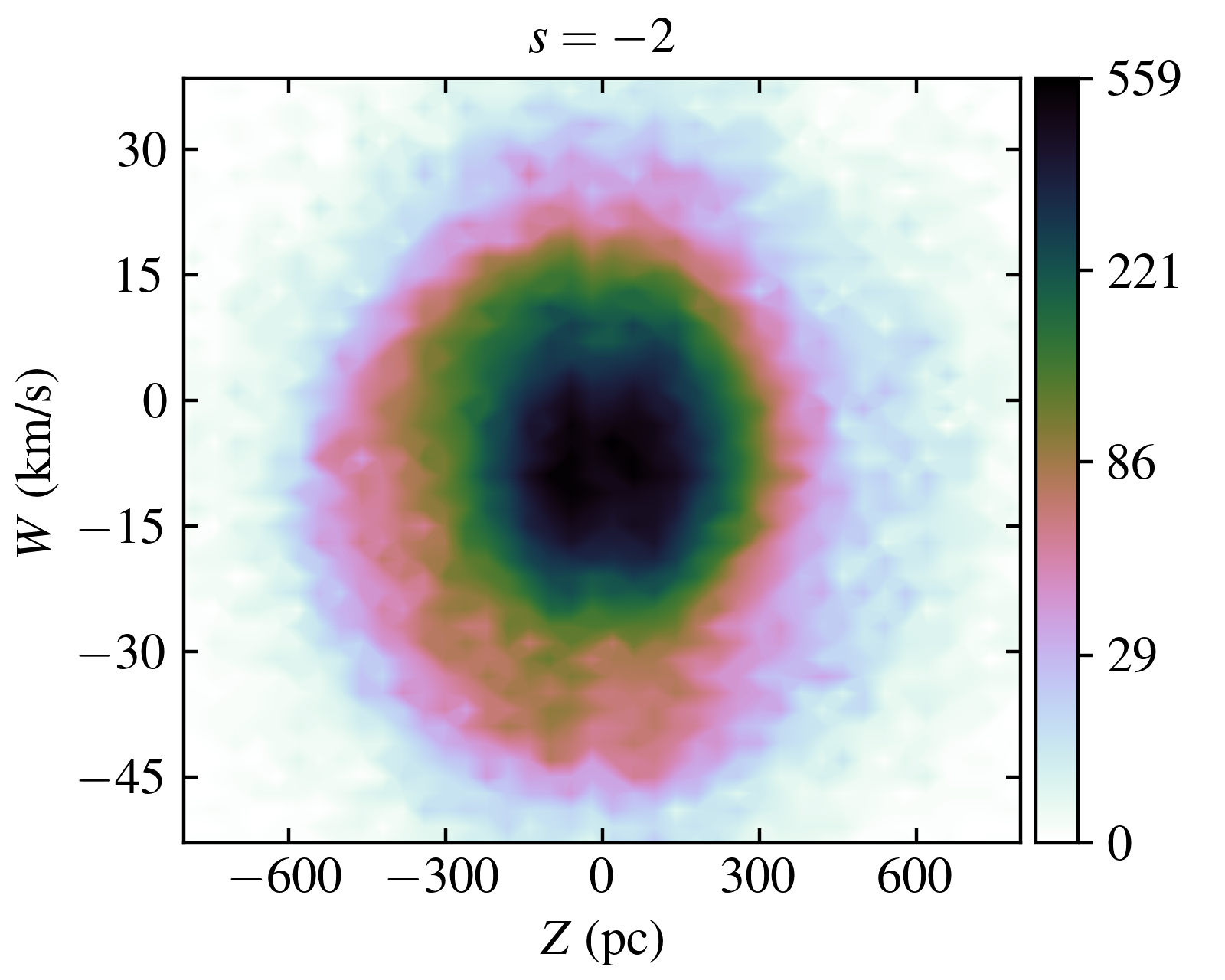}
\end{subfigure}
\begin{subfigure}{.5\textwidth}
    \centering
    \includegraphics[width=.95\linewidth]{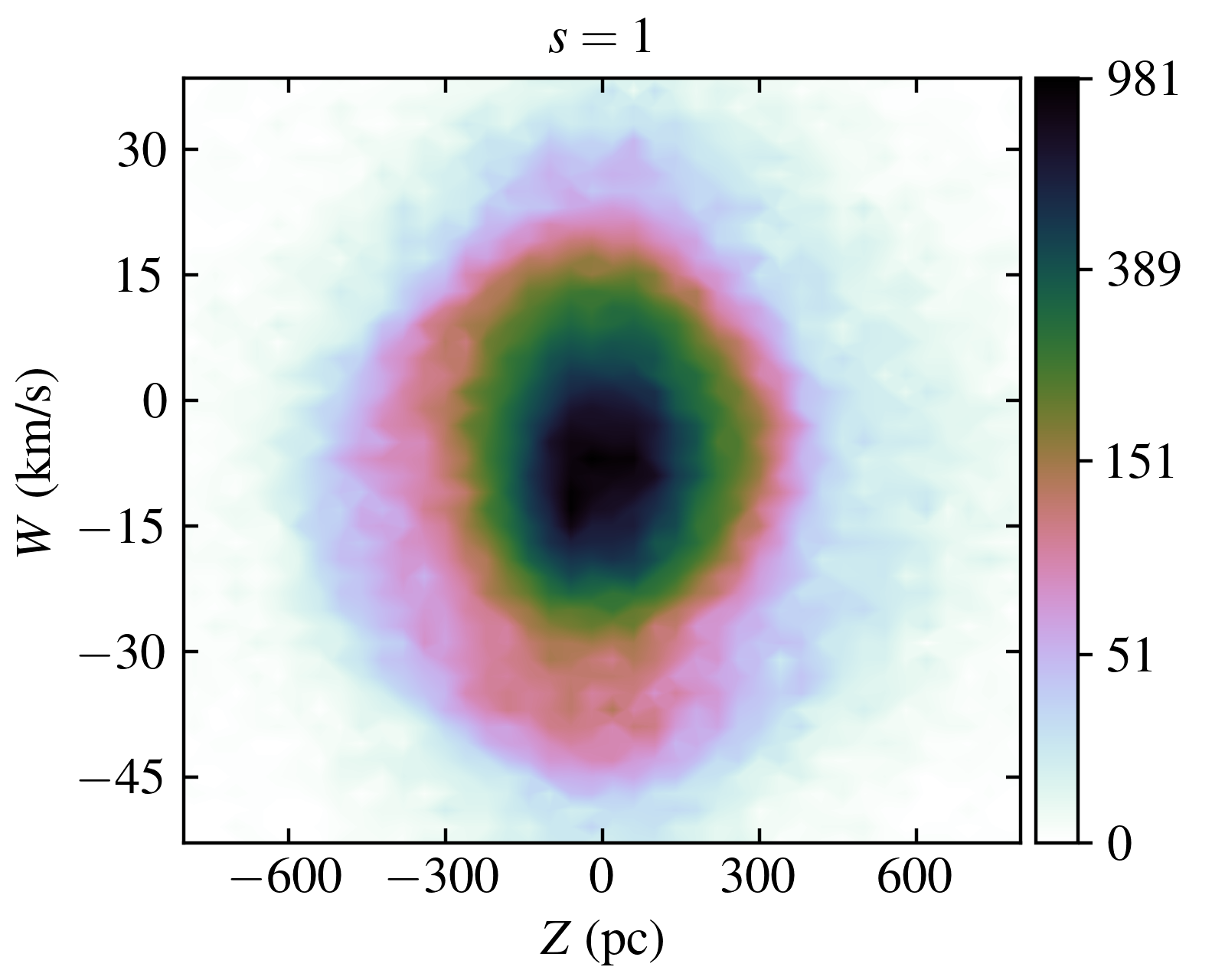}
\end{subfigure}
\par\bigskip\bigskip
\begin{subfigure}{.5\textwidth}
    \centering
    \includegraphics[width=.95\linewidth]{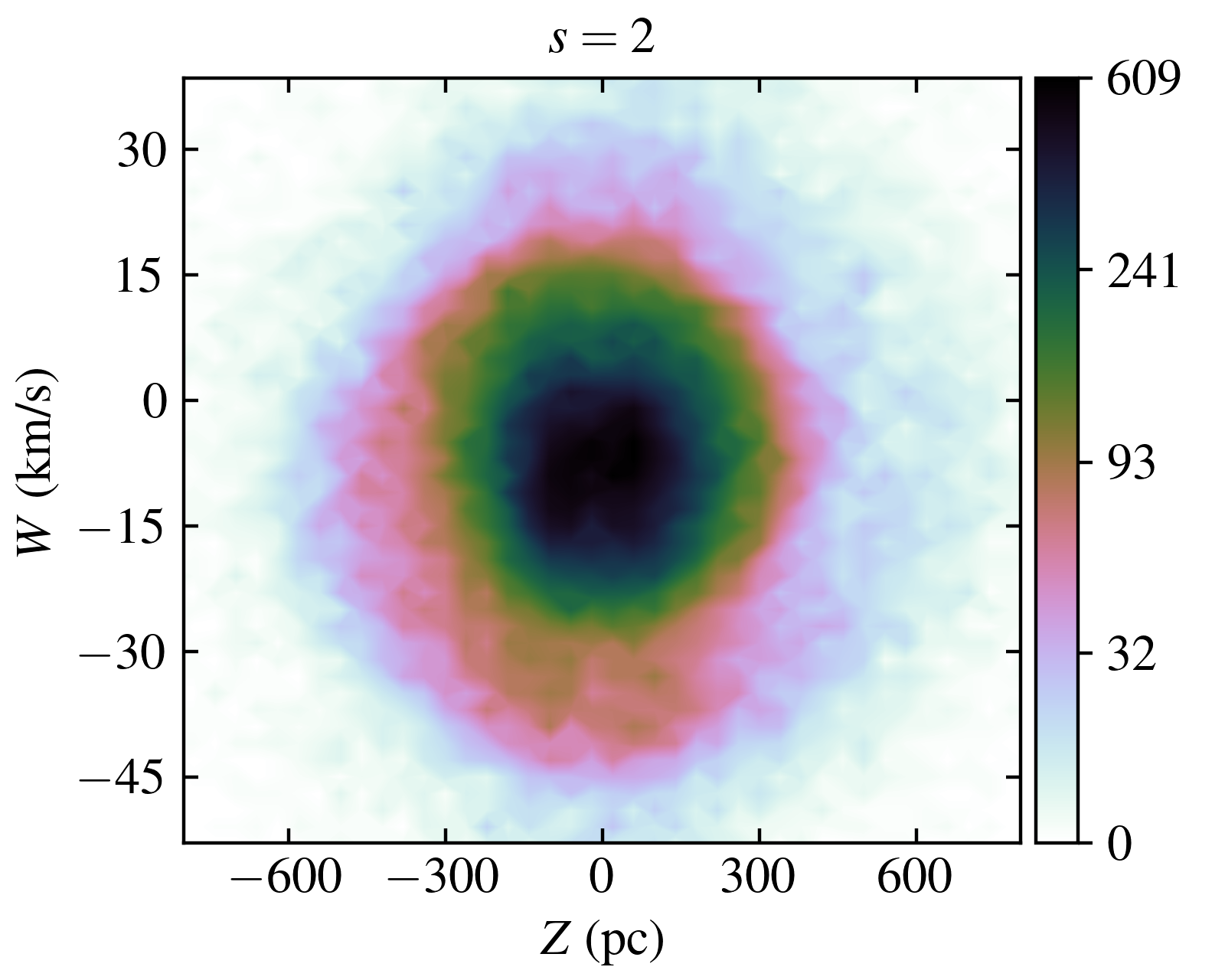}
\end{subfigure}
\begin{subfigure}{.5\textwidth}
    \centering
    \includegraphics[width=.95\linewidth]{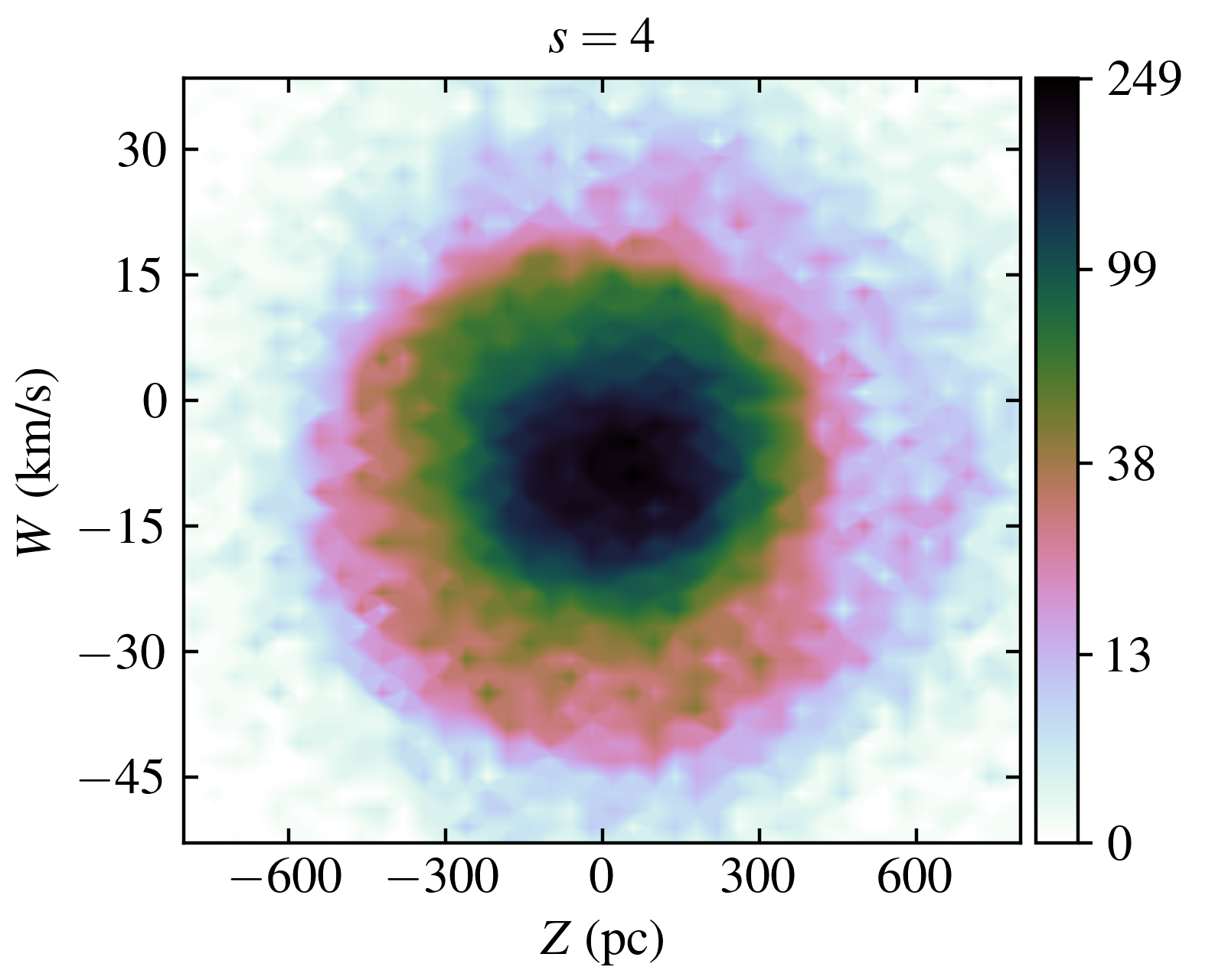}
\end{subfigure}
\caption{Same as Fig.~\ref{fig:hists}, but for data samples $s=\{-5,-3,-2,1,2,4,5\}$ and $\sqrt{X^2+Y^2}<300~\pc$.}
\label{fig:hists_app1}
\end{figure*}

\begin{figure*}
\ContinuedFloat
\begin{subfigure}{.5\textwidth}
    \centering
    \includegraphics[width=.95\linewidth]{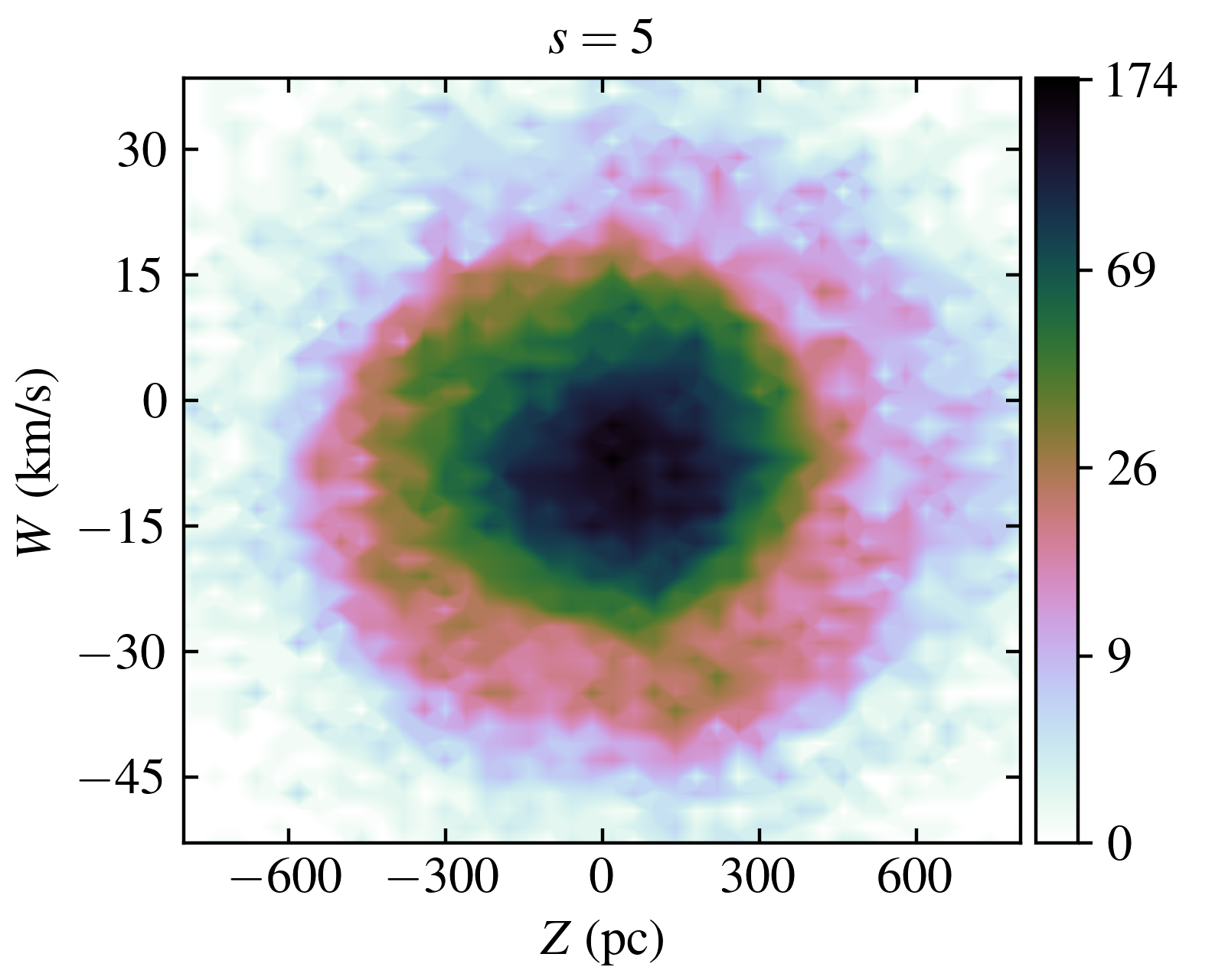}
\end{subfigure}
\begin{subfigure}{.5\textwidth}
    \centering
    \includegraphics[width=.95\linewidth]{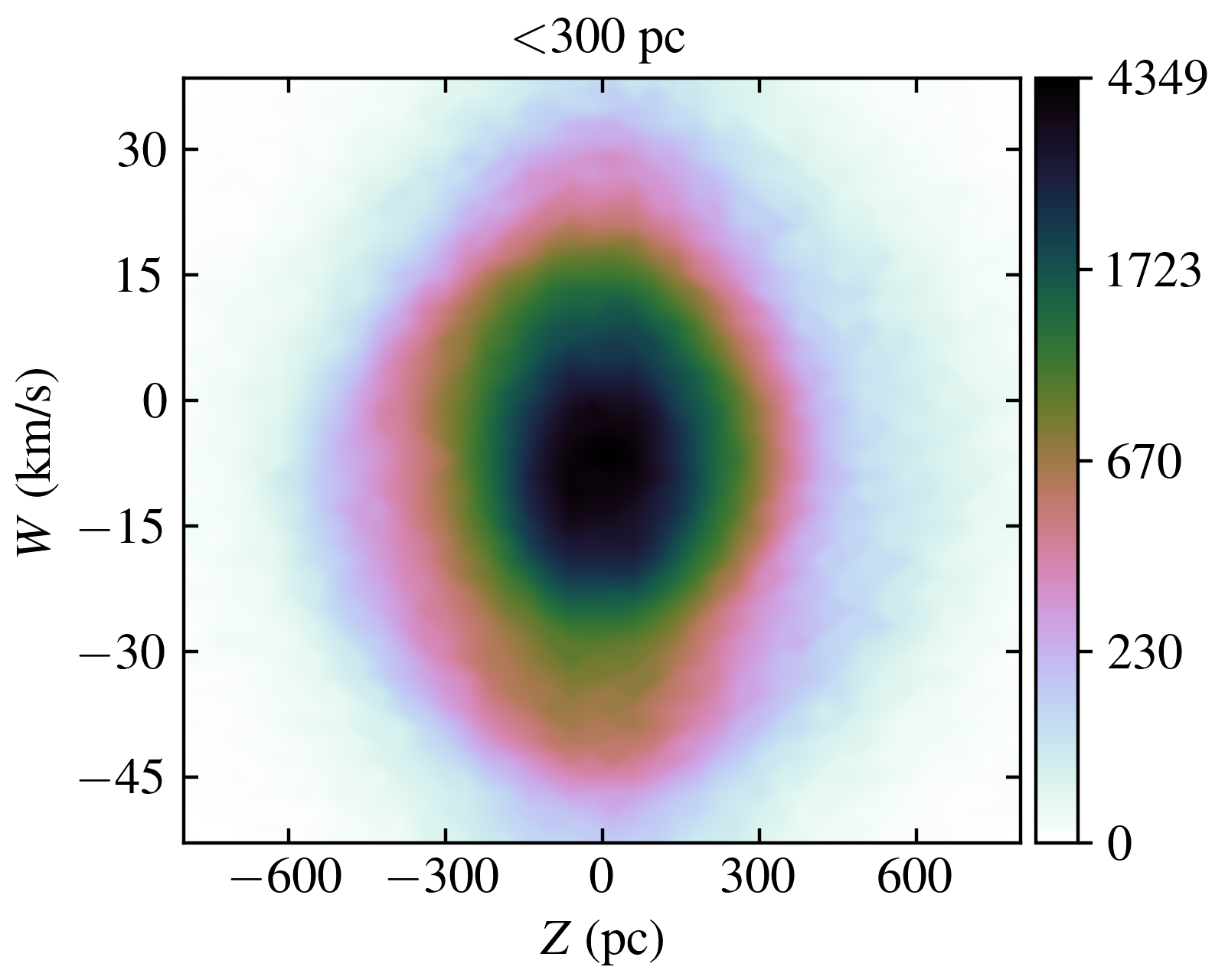}
\end{subfigure}
\caption{Continued.}
\end{figure*}

\begin{figure*}
\begin{subfigure}{.5\textwidth}
    \centering
    \includegraphics[width=1.\linewidth]{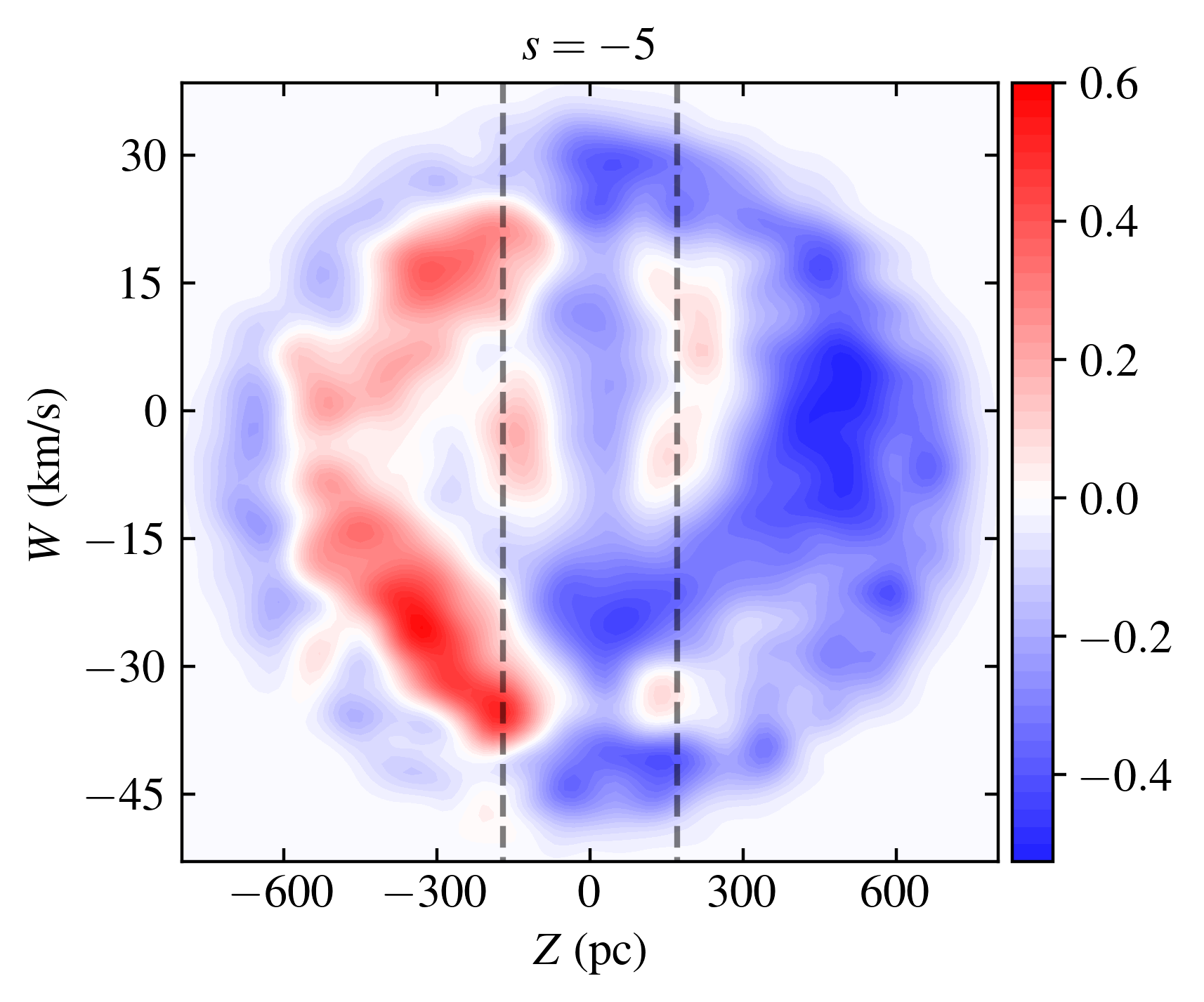}
\end{subfigure}
\begin{subfigure}{.5\textwidth}
    \centering
    \includegraphics[width=1.\linewidth]{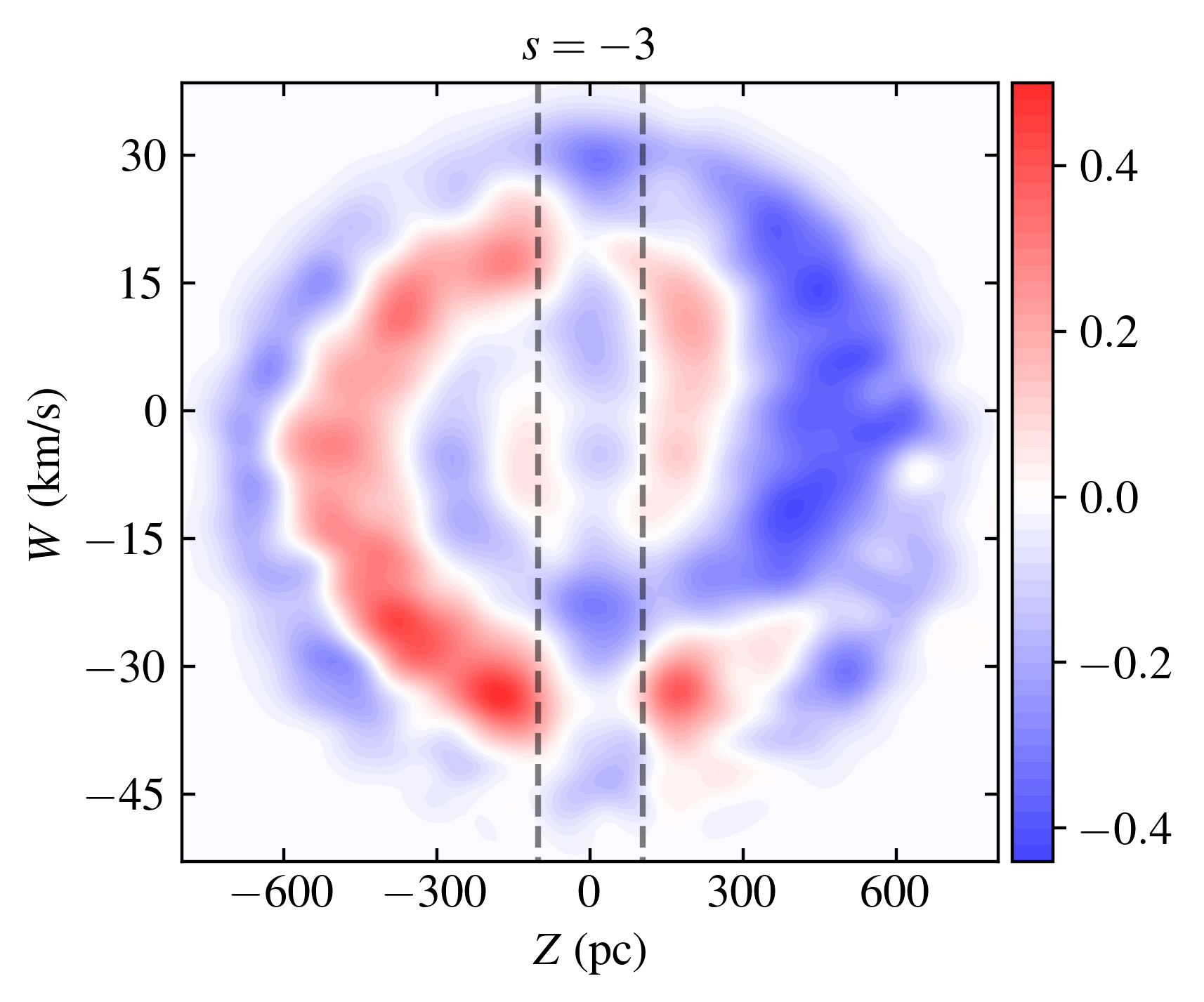}
\end{subfigure}
\caption{Same as Fig.~\ref{fig:spirals}, but for $s=\{-5,-3,-2,1,2,4,5\}$ and $\sqrt{X^2+Y^2}<300~\pc$.}
\label{fig:spirals_app1}
\end{figure*}

\begin{figure*}
\ContinuedFloat
\begin{subfigure}{.5\textwidth}
    \centering
    \includegraphics[width=1.\linewidth]{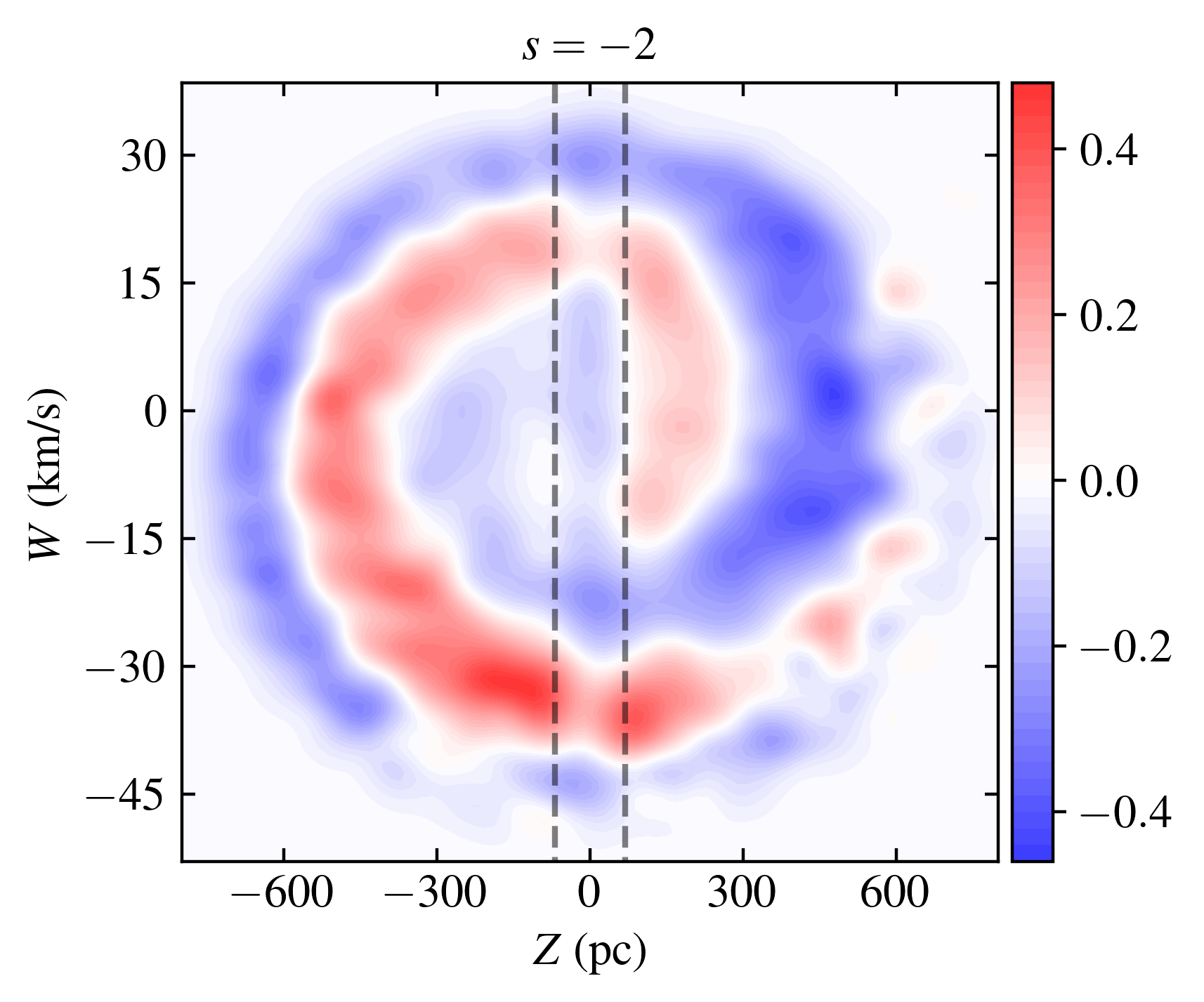}
\end{subfigure}
\begin{subfigure}{.5\textwidth}
    \centering
    \includegraphics[width=1.\linewidth]{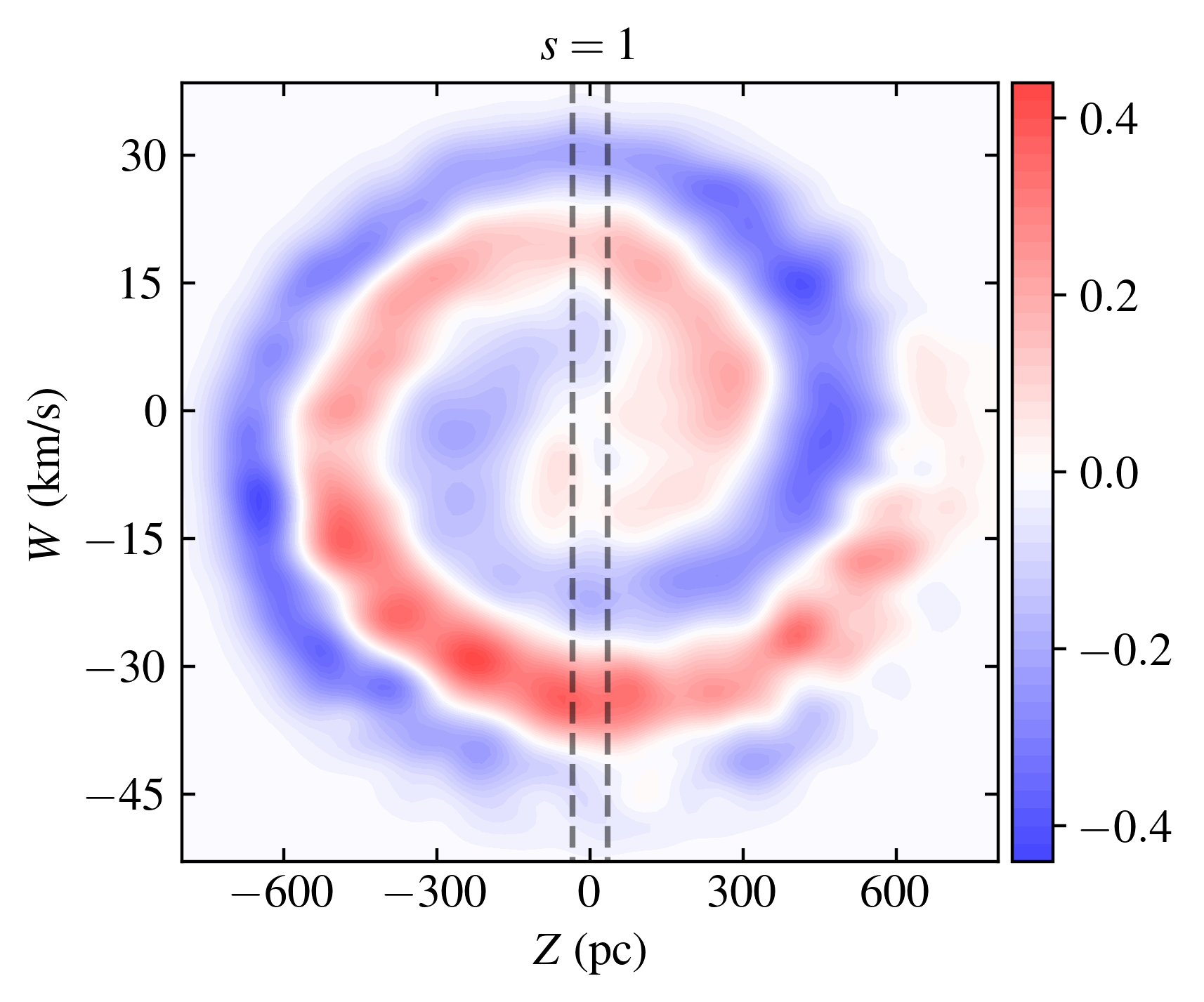}
\end{subfigure}
\par\bigskip
\begin{subfigure}{.5\textwidth}
    \centering
    \includegraphics[width=1.\linewidth]{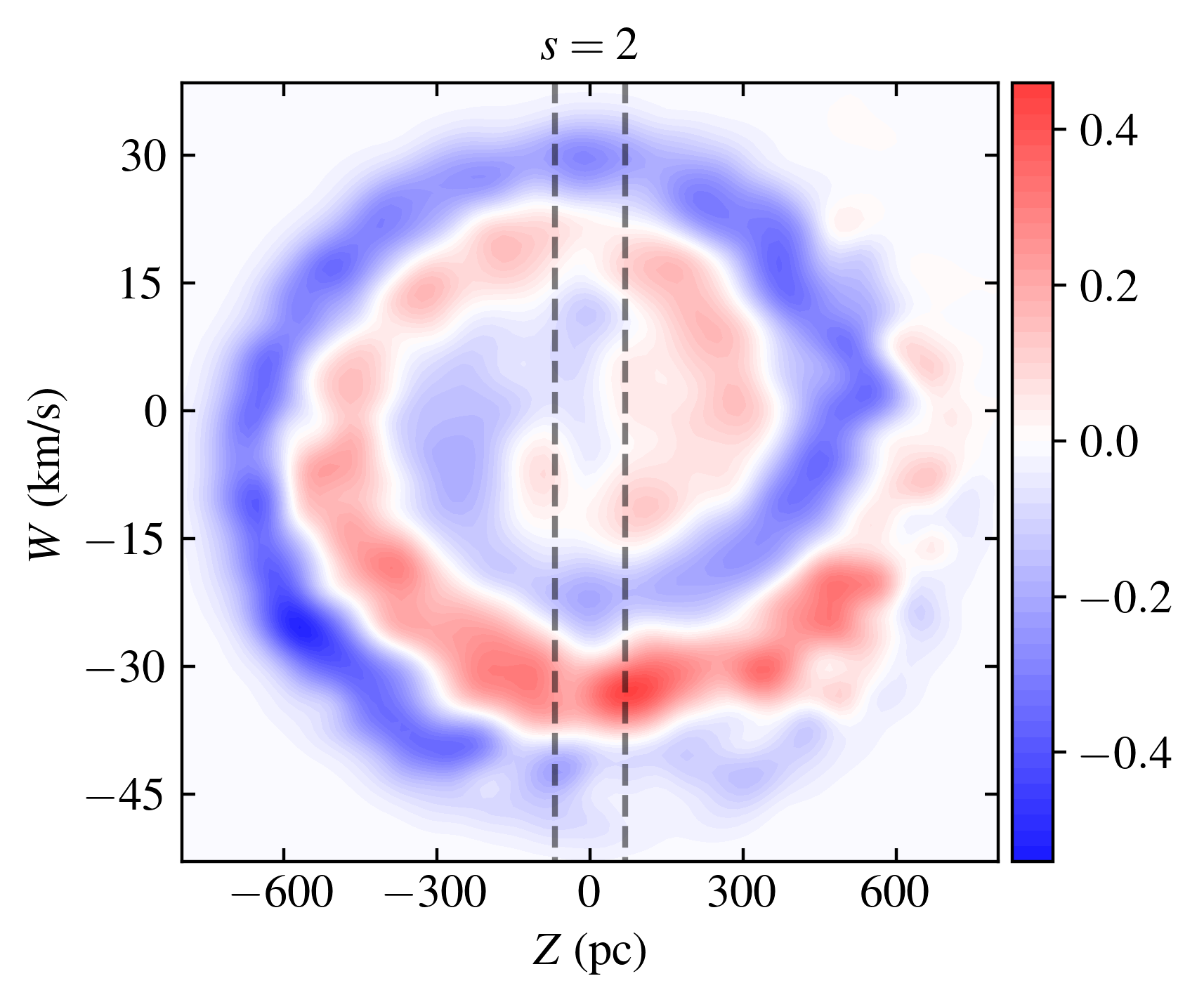}
\end{subfigure}
\begin{subfigure}{.5\textwidth}
    \centering
    \includegraphics[width=1.\linewidth]{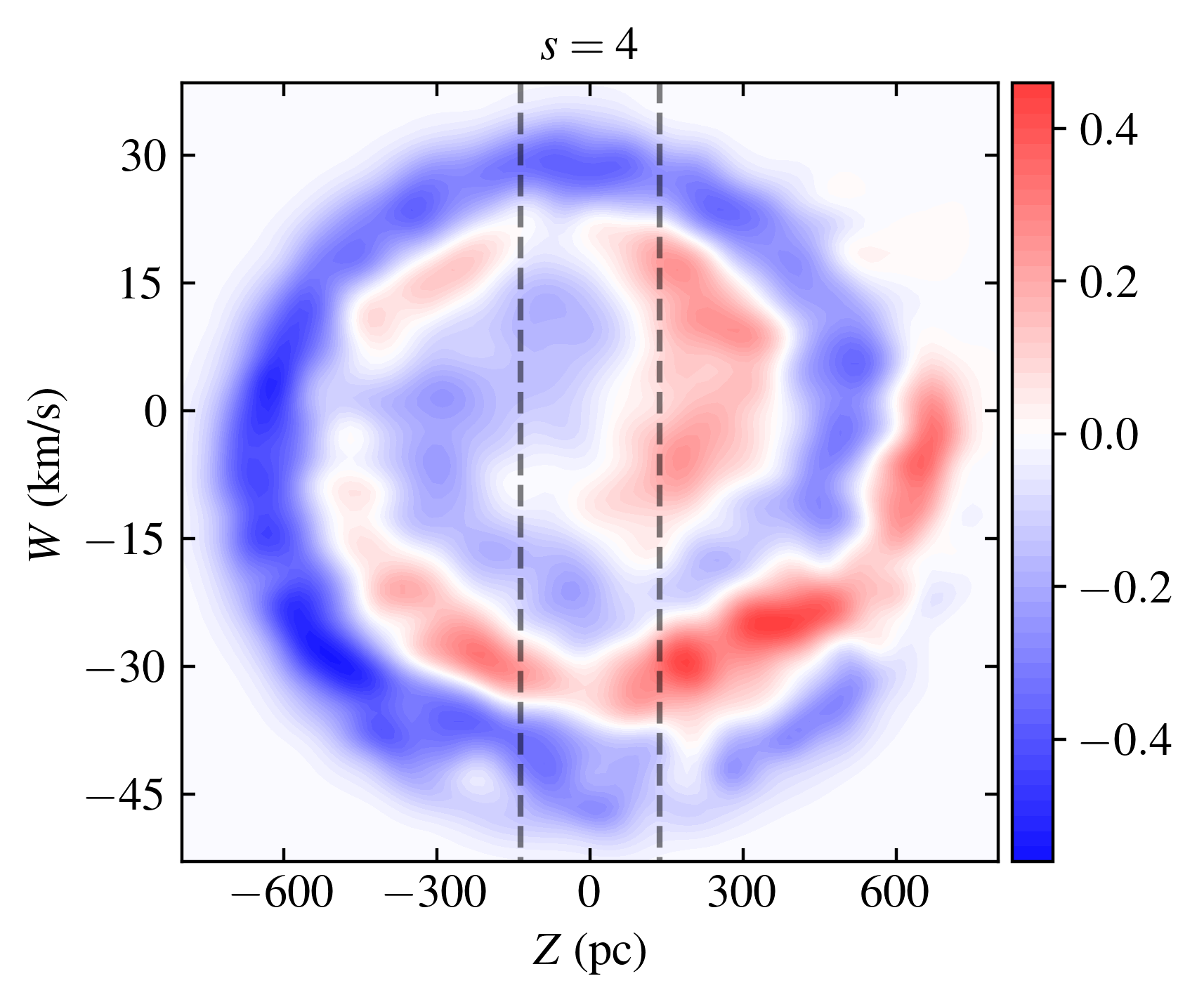}
\end{subfigure}
\par\bigskip
\begin{subfigure}{.5\textwidth}
    \centering
    \includegraphics[width=1.\linewidth]{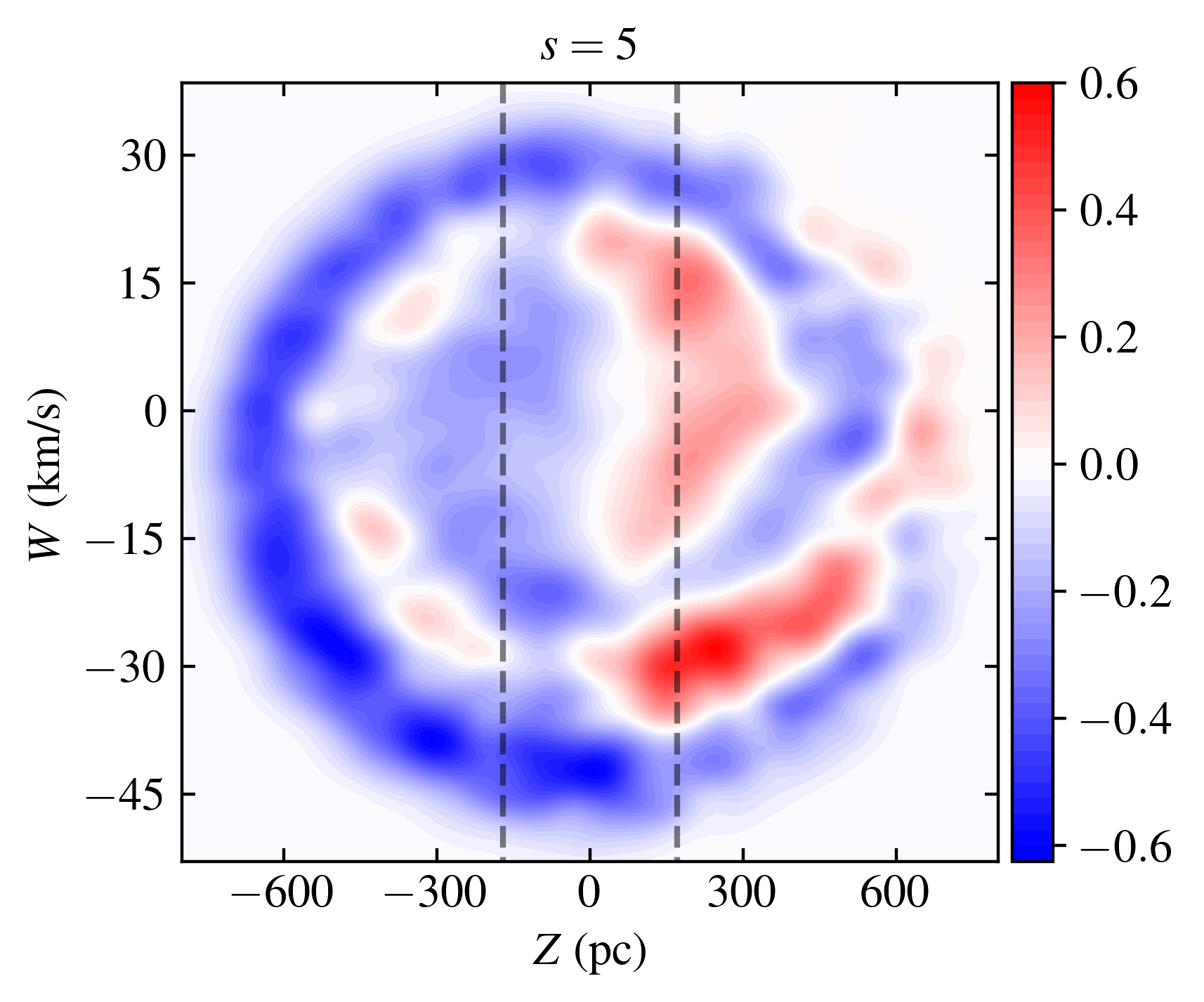}
\end{subfigure}
\begin{subfigure}{.5\textwidth}
    \centering
    \includegraphics[width=1.\linewidth]{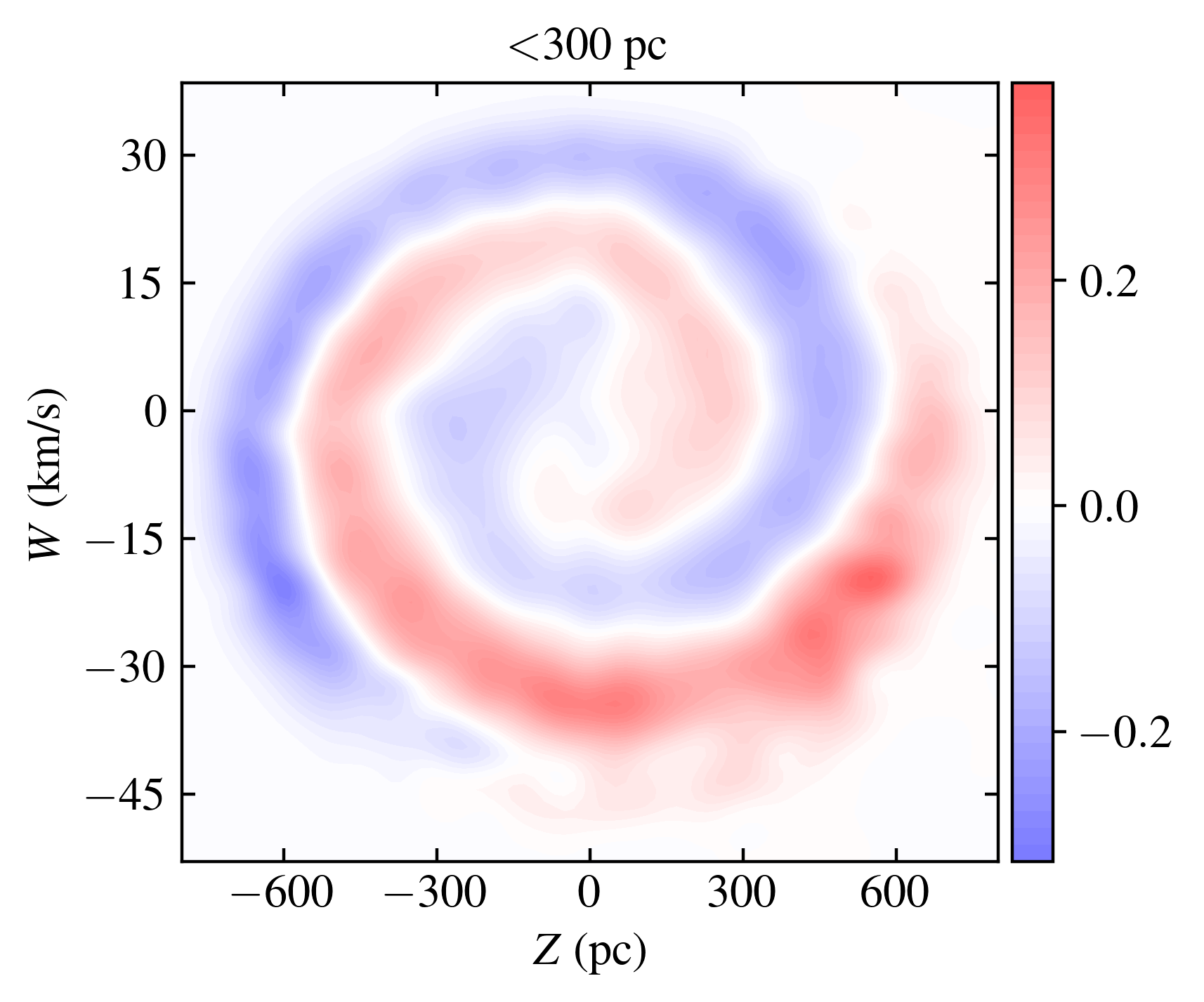}
\end{subfigure}
\caption{Continued}
\end{figure*}

\begin{figure*}
\begin{subfigure}{1.\textwidth}
    \centering
    \includegraphics[width=1.\linewidth]{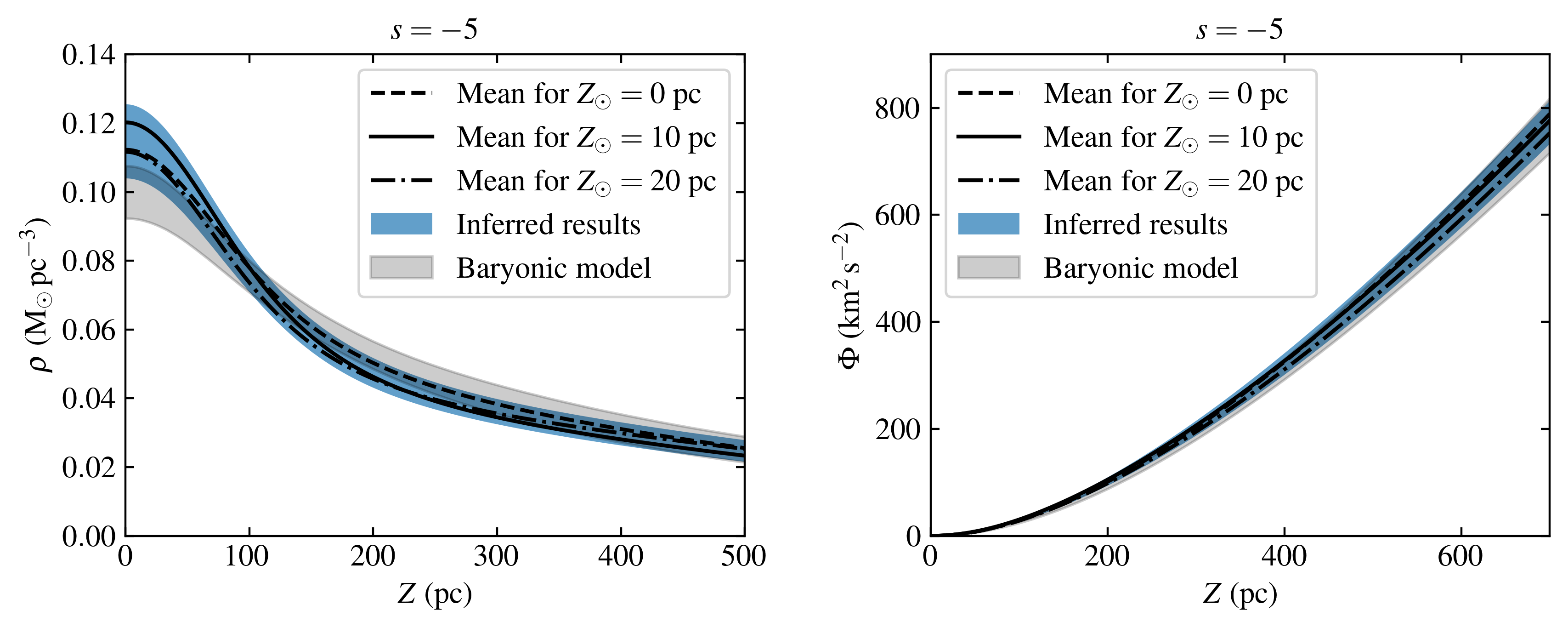}
\end{subfigure}
\par\bigskip
\begin{subfigure}{1.\textwidth}
    \centering
    \includegraphics[width=1.\linewidth]{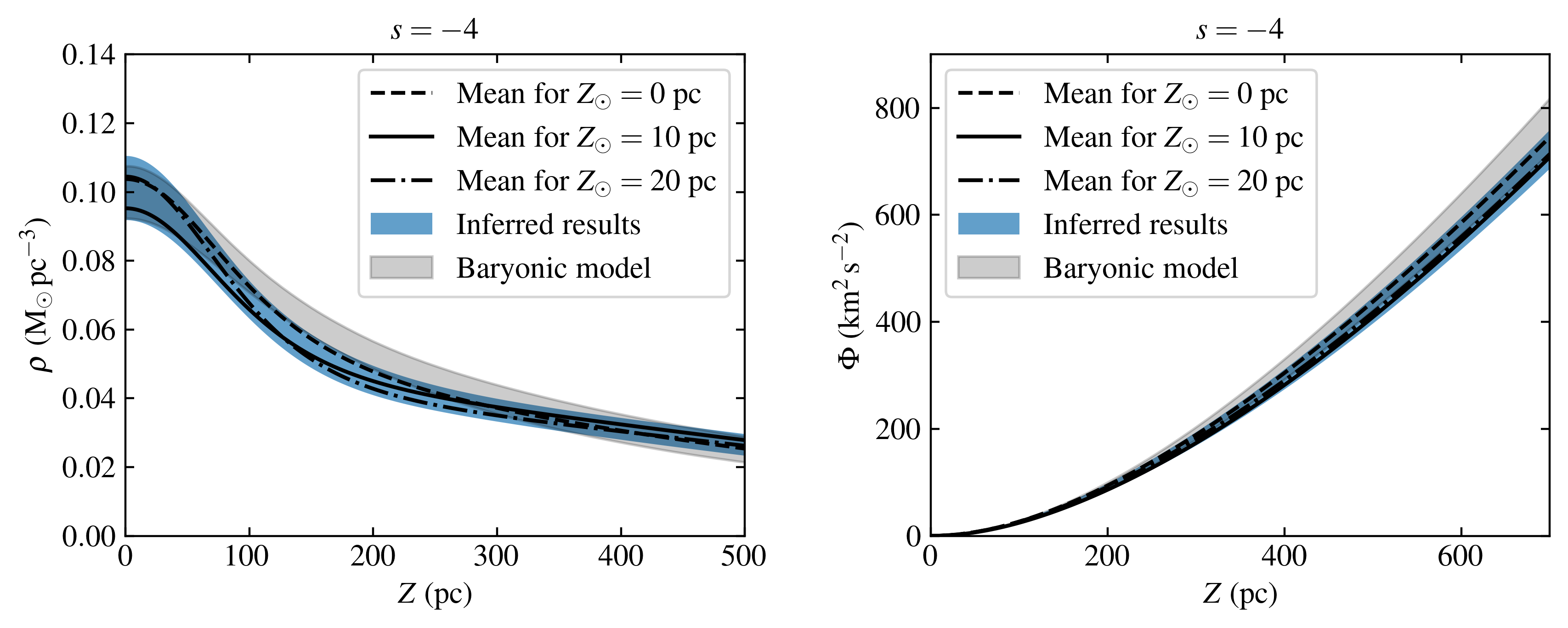}
\end{subfigure}
\par\bigskip
\begin{subfigure}{1.\textwidth}
    \centering
    \includegraphics[width=1.\linewidth]{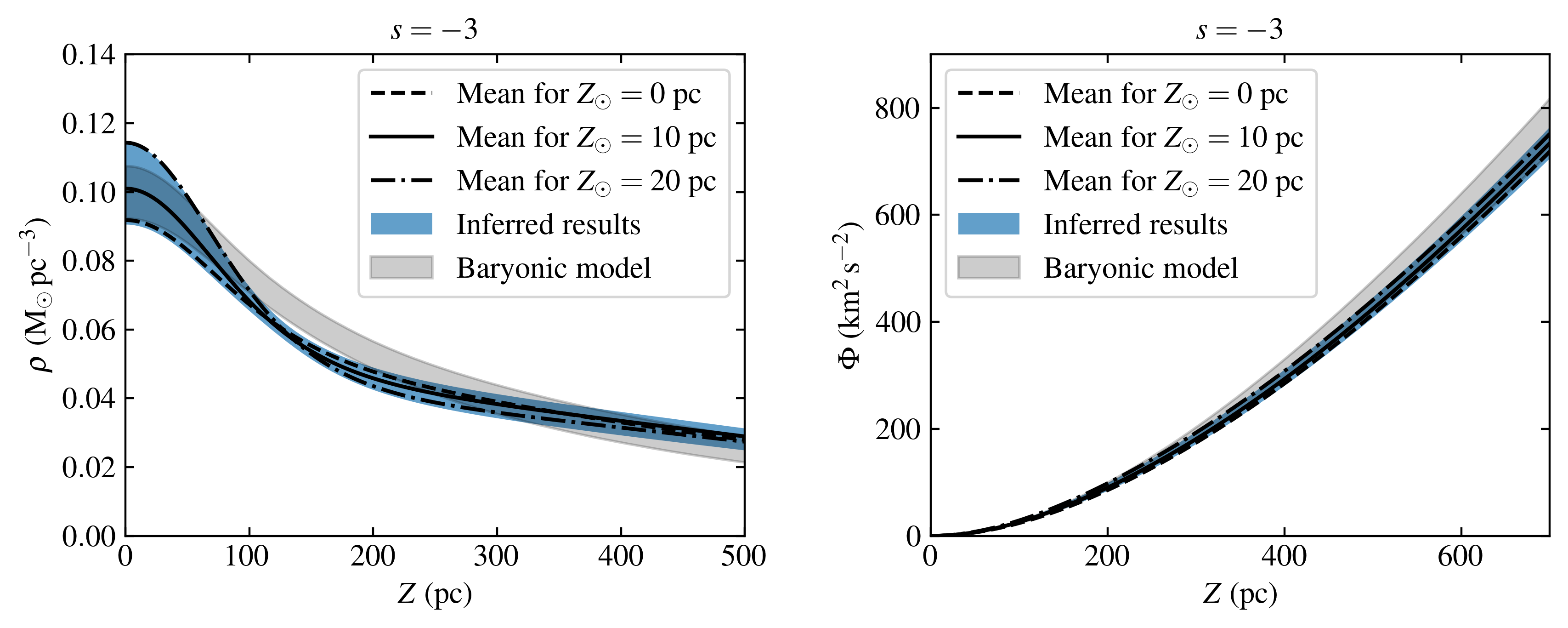}
\end{subfigure}
\caption{Same as Fig.~\ref{fig:jks}, but for $s=\{-5,-4,-3,-1,1,3,4,5\}$ and $\sqrt{X^2+Y^2}<300~\pc$.}
\label{fig:jks_app1}
\end{figure*}
\begin{figure*}
\ContinuedFloat
\begin{subfigure}{1.\textwidth}
    \centering
    \includegraphics[width=1.\linewidth]{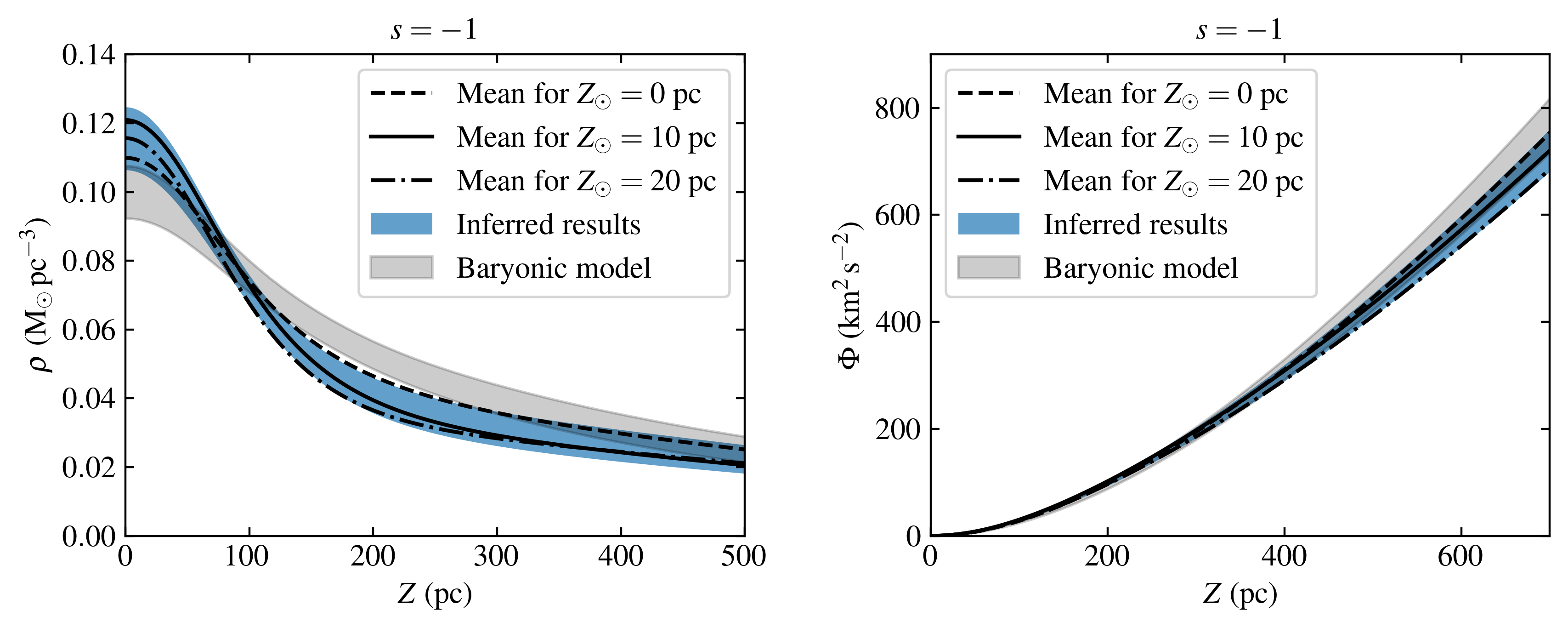}
\end{subfigure}
\par\bigskip
\begin{subfigure}{1.\textwidth}
    \centering
    \includegraphics[width=1.\linewidth]{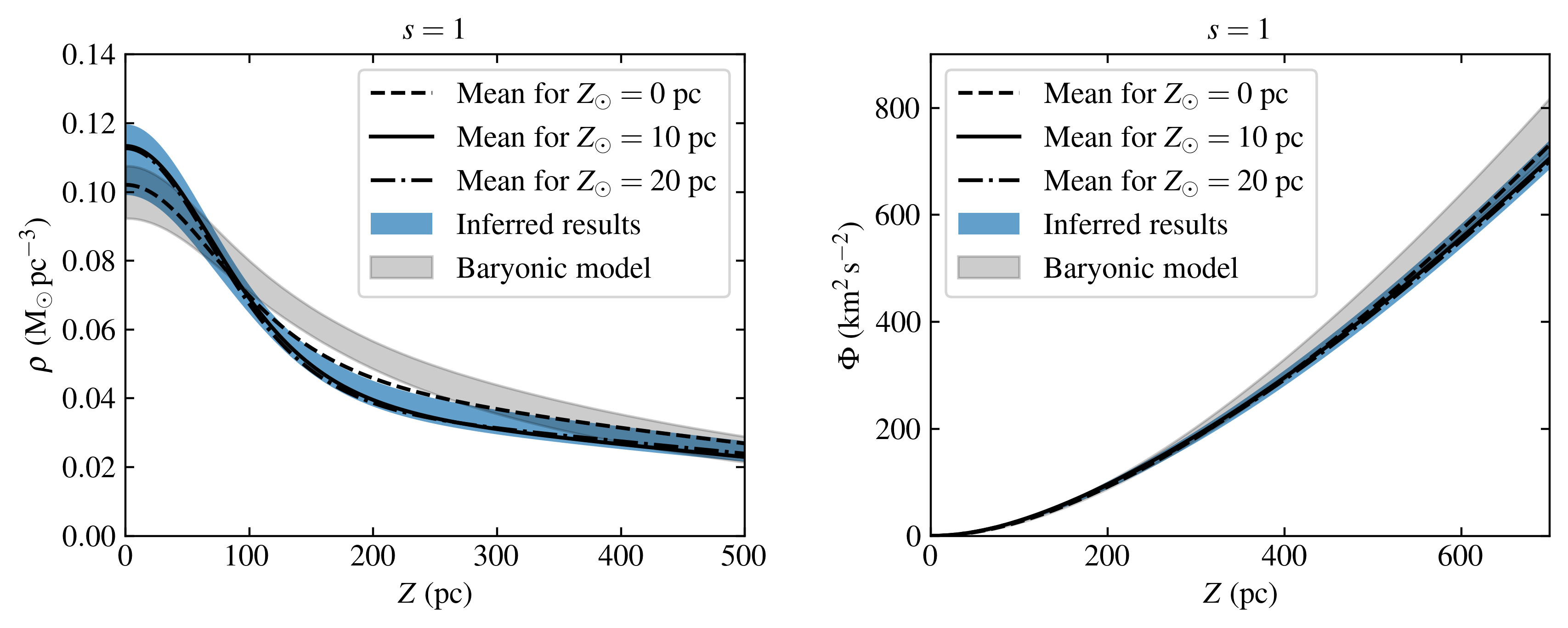}
\end{subfigure}
\par\bigskip
\begin{subfigure}{1.\textwidth}
    \centering
    \includegraphics[width=1.\linewidth]{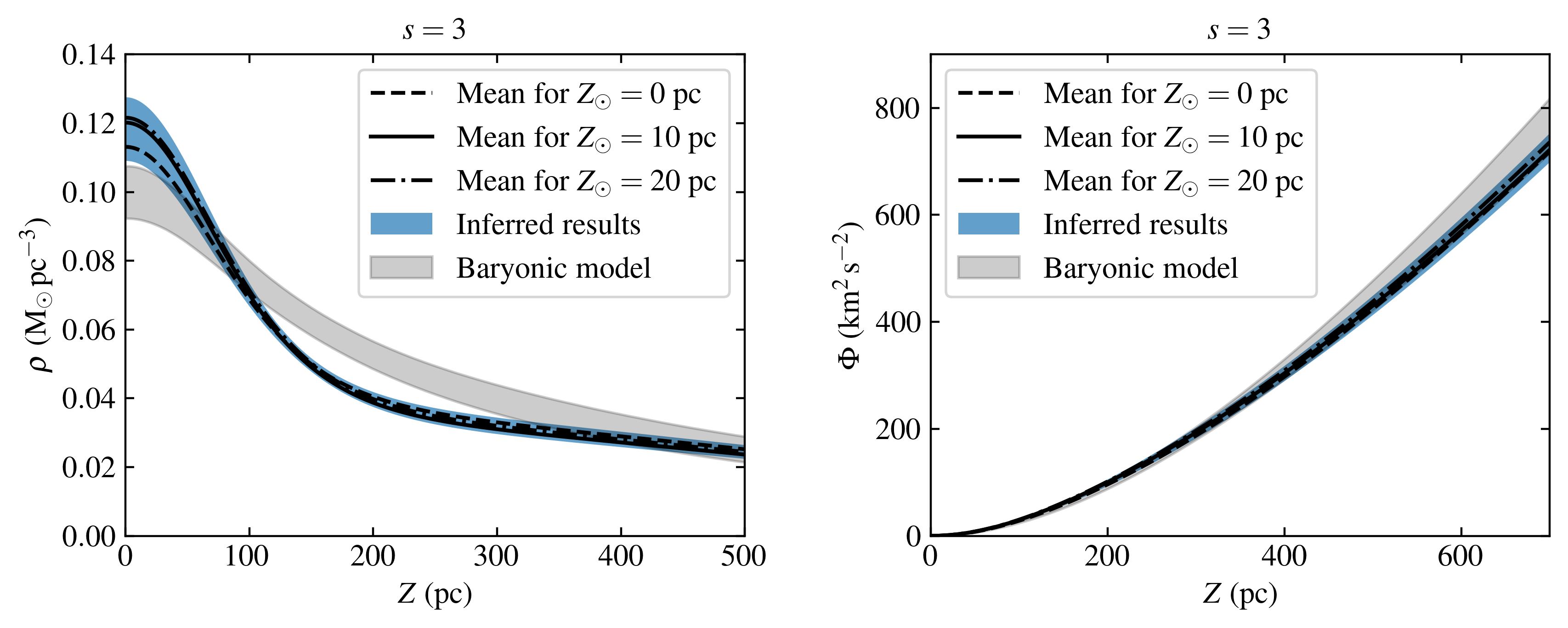}
\end{subfigure}
\caption{Continued.}
\end{figure*}
\begin{figure*}
\ContinuedFloat
\begin{subfigure}{1.\textwidth}
    \centering
    \includegraphics[width=1.\linewidth]{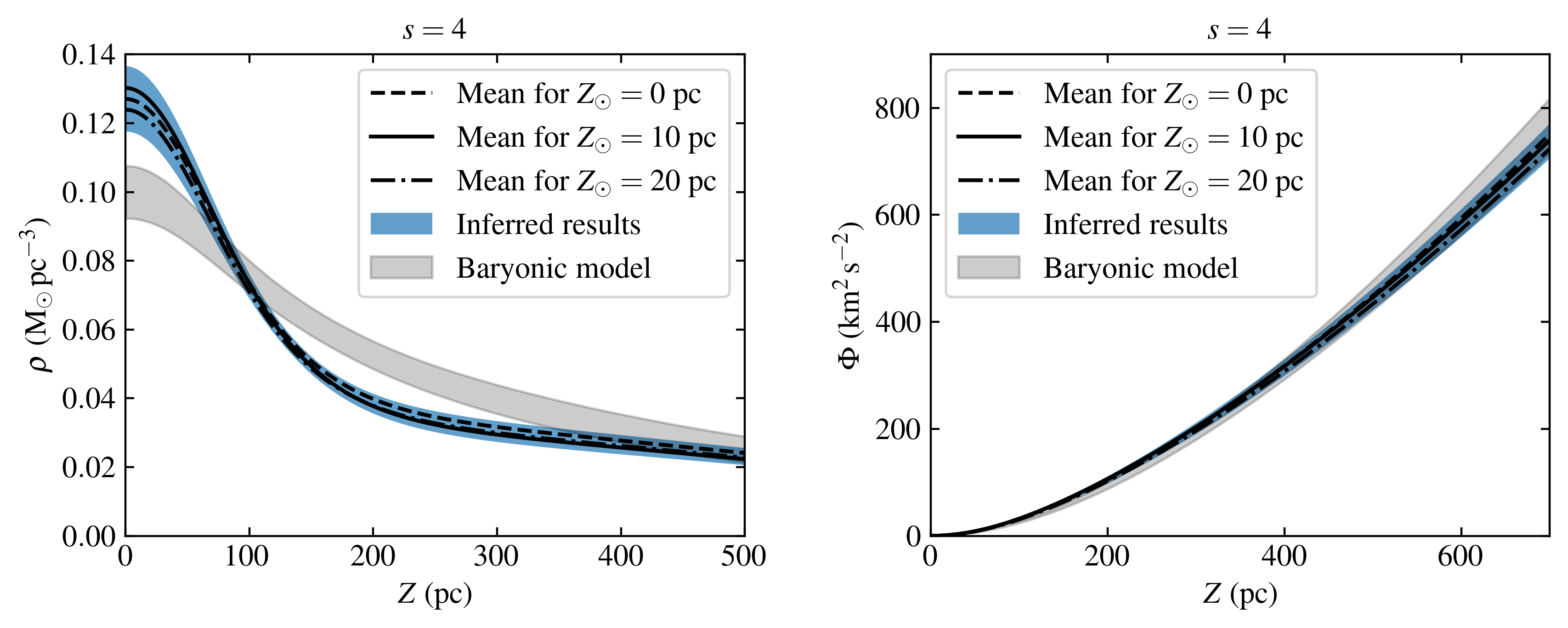}
\end{subfigure}
\par\bigskip
\begin{subfigure}{1.\textwidth}
    \centering
    \includegraphics[width=1.\linewidth]{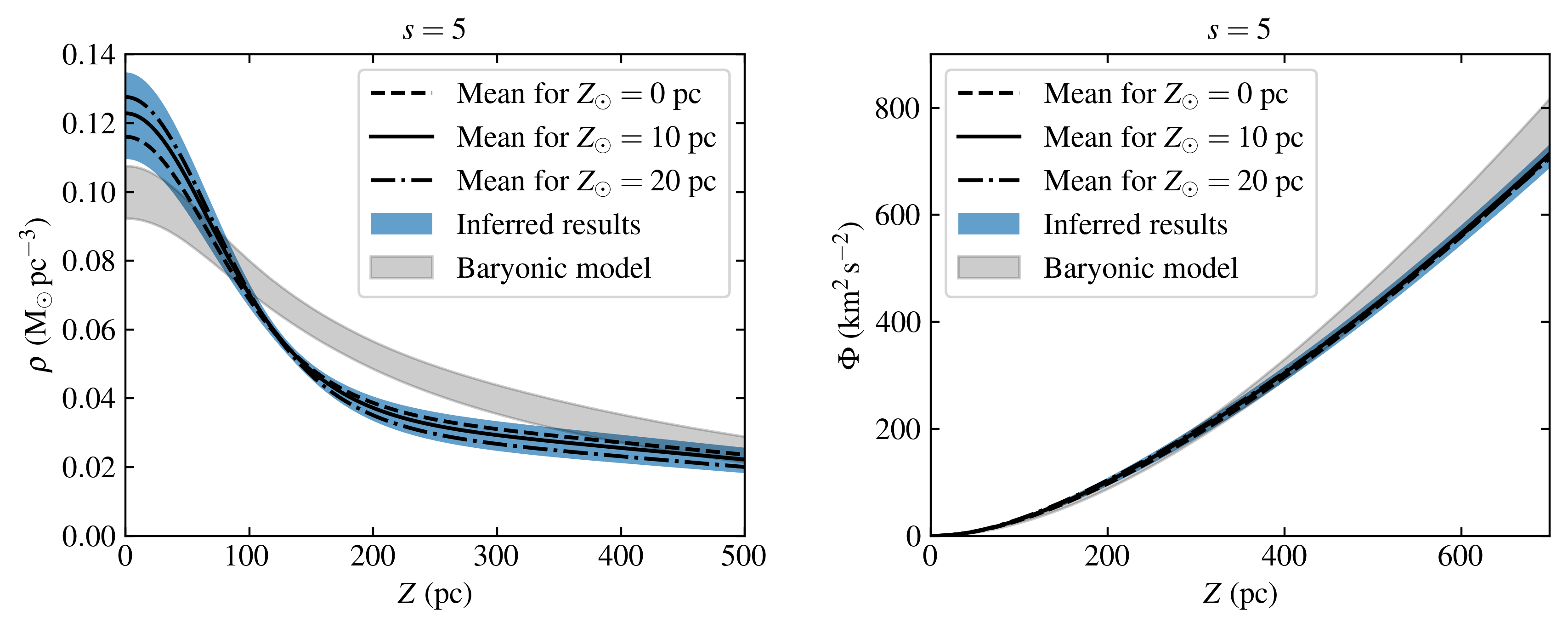}
\end{subfigure}
\par\bigskip
\begin{subfigure}{1.\textwidth}
    \centering
    \includegraphics[width=1.\linewidth]{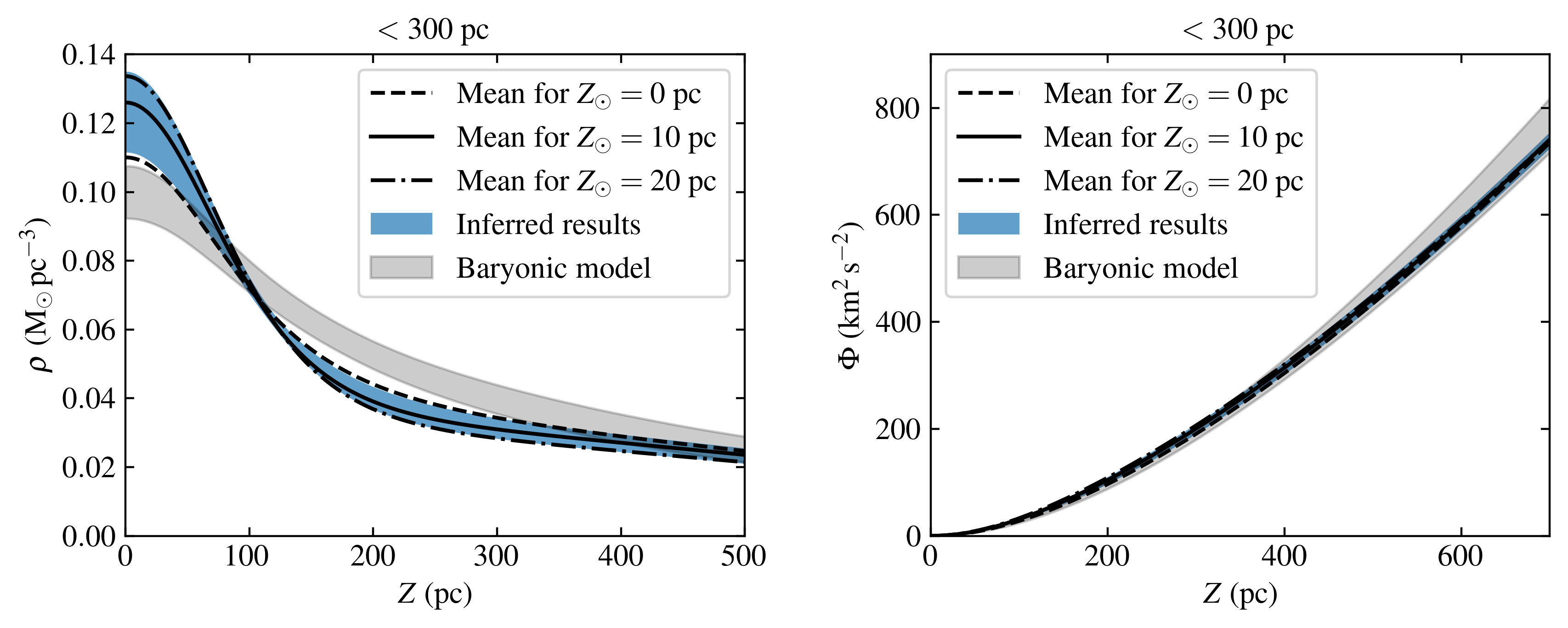}
\end{subfigure}
\caption{Continued.}
\end{figure*}

\end{appendix}

\end{document}